\documentclass[aps,prl,floatfix,twocolumn,superscriptaddress,tightenlines,nobibnotes,amsmath,amssymb]{revtex4-1}

\usepackage{graphicx}
\usepackage{amsmath}
\usepackage{amssymb}
\usepackage{epstopdf}
\usepackage{color}
\usepackage{natbib}
\usepackage{upgreek}

\newcommand{\bra}[1] {\langle #1\rvert}
\newcommand{\ket}[1] {\lvert#1 \rangle}

\newcommand*{\rom}[1]{\expandafter\@slowromancap\romannumeral #1@}

\begin{document}

\title{Two-Electron Spin Correlations in Precision Placed Donors in Silicon}
\author{M. A. Broome}
\altaffiliation{These authors contributed equally to this work}
\affiliation{Centre of Excellence for Quantum Computation and Communication Technology, School of Physics, University of New South Wales, Sydney, New South Wales 2052, Australia}
\author{S. K. Gorman} 
\altaffiliation{These authors contributed equally to this work}
\affiliation{Centre of Excellence for Quantum Computation and Communication Technology, School of Physics, University of New South Wales, Sydney, New South Wales 2052, Australia}
\author{M. G. House} 
\affiliation{Centre of Excellence for Quantum Computation and Communication Technology, School of Physics, University of New South Wales, Sydney, New South Wales 2052, Australia}
\author{S. J. Hile} 
\affiliation{Centre of Excellence for Quantum Computation and Communication Technology, School of Physics, University of New South Wales, Sydney, New South Wales 2052, Australia}
\author{J. G. Keizer} 
\affiliation{Centre of Excellence for Quantum Computation and Communication Technology, School of Physics, University of New South Wales, Sydney, New South Wales 2052, Australia}
\author{D. Keith} 
\affiliation{Centre of Excellence for Quantum Computation and Communication Technology, School of Physics, University of New South Wales, Sydney, New South Wales 2052, Australia}
\author{\\C. D. Hill} 
\affiliation{Centre of Excellence for Quantum Computation and Communication Technology, School of Physics, The University of Melbourne, Parkville, Victoria 3010, Australia}
\author{T. F. Watson} 
\affiliation{Centre of Excellence for Quantum Computation and Communication Technology, School of Physics, University of New South Wales, Sydney, New South Wales 2052, Australia}
\author{W. J. Baker}
\affiliation{Centre of Excellence for Quantum Computation and Communication Technology, School of Physics, University of New South Wales, Sydney, New South Wales 2052, Australia}
\author{L. C. L. Hollenberg} 
\affiliation{Centre of Excellence for Quantum Computation and Communication Technology, School of Physics, The University of Melbourne, Parkville, Victoria 3010, Australia}
\author{M. Y. Simmons}
\email{michelle.simmons@unsw.edu.au}
\affiliation{Centre of Excellence for Quantum Computation and Communication Technology, School of Physics, University of New South Wales, Sydney, New South Wales 2052, Australia}
\date{\today}

\begin{abstract}
\section{Abstract}
Substitutional donor atoms in silicon are promising qubits for quantum computation with extremely long relaxation and dephasing times demonstrated. One of the critical challenges of scaling these systems is determining inter-donor distances to achieve controllable wavefunction overlap while at the same time performing high fidelity spin readout on each qubit. Here we achieve such a device by means of scanning tunnelling microscopy lithography. We measure anti-correlated spin states between two donor-based spin qubits in silicon separated by 16${\pm}1$~nm. By utilizing an asymmetric system with two phosphorus donors at one qubit site and one on the other (2P-1P), we demonstrate that the exchange interaction can be turned on and off via electrical control of two in-plane phosphorus doped detuning gates. We determine the tunnel coupling between the 2P-1P system to be 200~MHz and provide a roadmap for the observation of two-electron coherent exchange oscillations.

\end{abstract}

\maketitle

\section{Introduction}

\begin{figure*}
\begin{center}
\includegraphics[width=0.8\textwidth]{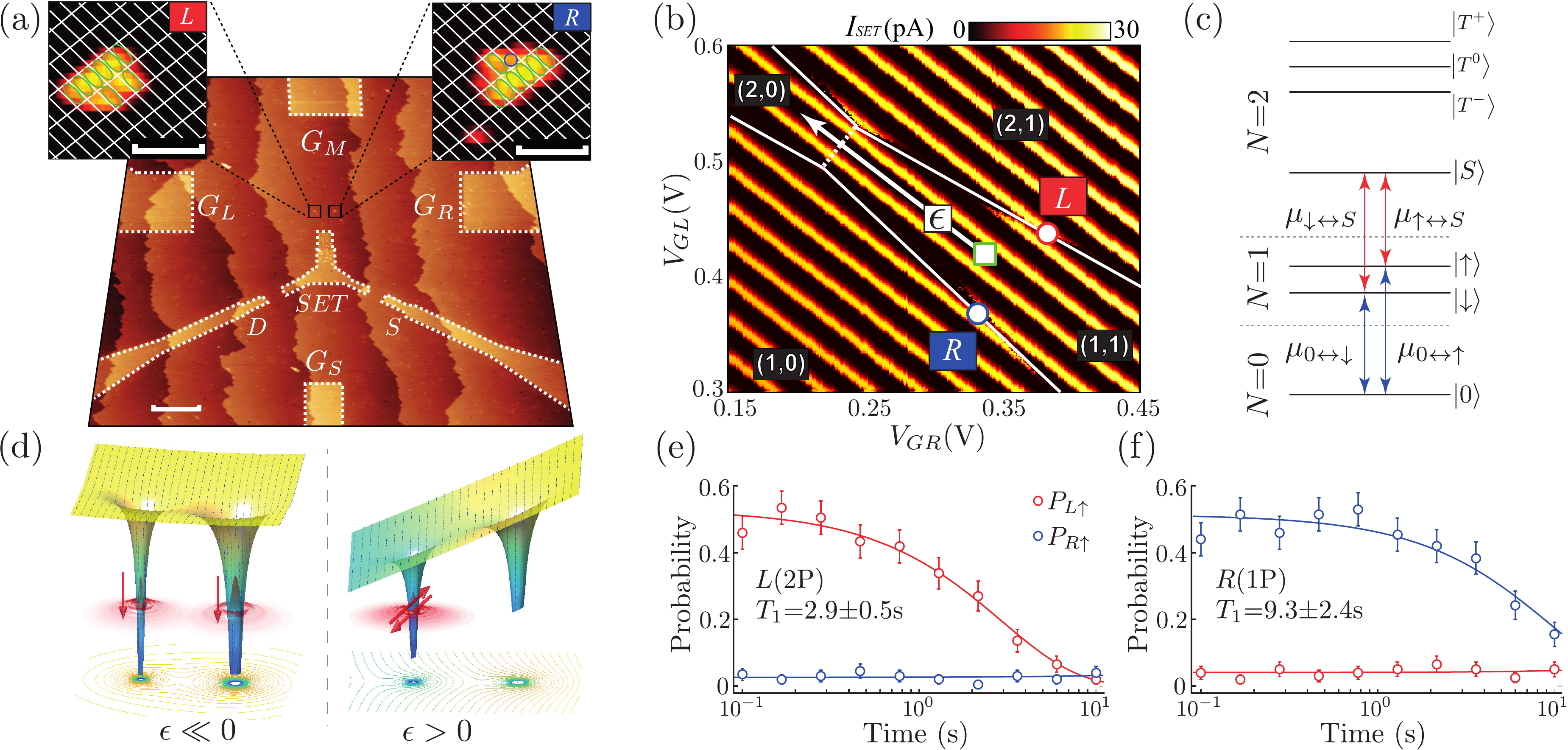}
\end{center}
\vspace{-0.5cm}
\caption{{ \bf Two qubit 2P-1P device with independent sequential readout.} {\bf a}, An STM micrograph of a precision placed two spin-qubit donor device showing the lighter coloured lithographic outline where the hydrogen mask has been removed. Two spin-qubits, L and R are separated by $16{\pm}1$~nm and sit equidistant at $19{\pm}1$~nm away from a larger readout structure which serves as both an electron reservoir and single-electron-transistor (SET) charge sensor with source (S) and drain (D) reservoirs and gates \{G\textsubscript{L}, G\textsubscript{M},  G\textsubscript{R}, G\textsubscript{S}\}, the scale bar is 20~nm. The insets show close-up STM micrographs of L and R where the green (blue) circles show fully (half) desorbed silicon dimers. White lines indicate the silicon dimer rows and the scale bars are 2~nm. {\bf b}, Current through the SET charge sensor as a function of $V_{\textrm{GL}}$ and $V_{\textrm{GR}}$ at the (1,1)-(2,0) charge transition. Electron spin readout is performed at the SET breaks (solid white lines, where tunnelling of qubit electrons to or from SET can occur) at the red and blue circles for L and R respectively. The approximate wait position for spin relaxation measurements is shown by the green square and the detuning axis between (1,1)-(2,0), $\epsilon$, is indicated by the white arrow. The dashed white line indicates where electrons can tunnel between qubit sites, i.e. where $\epsilon{=}0$. {\bf c}, The relevant electrochemical potentials in a magnetic field for spin readout of L (red arrows) and R (blue arrows). {\bf d}, A schematic representation of the controllable exchange interaction in a 2P-1P donor spin-qubit system. For detuning $\epsilon{\ll}0$ the electrons are in the (1,1) charge configuration and the spins are independent. For $\epsilon{>}0$ the ground state (2,0) charge configuration is the two-electron singlet state. {\bf e}, and {\bf f}, Independent spin readout of L (R) demonstrated by spin relaxation, when the electron on R (L) is deterministically loaded with $\ket{\downarrow}$. In each case the qubit initially prepared as spin down shows no decay behaviour indicating that the readout is independent at this detuning position, i.e. the exchange is negligible at the readout positions. All measurements were performed with $B_z{=}2.5$~T.}
\vspace{-0.5cm}
\label{fig:readout}
\end{figure*}

Controlling the interaction strength between two quantum particles lies at the heart of quantum information processing. One must have access to classical control fields that, whilst tuning the environment of quantum particles, are sufficiently decoupled from them as to not disturb their quantum states~\cite{nielsen2010quantum}. Physical systems ranging from trapped ions~\cite{haffner2008}, single photons~\cite{UniversalLO:OBrien}, superconducting circuits~\cite{clarke2008} and semiconductor quantum dots~\cite{petta2005} have demonstrated this exquisite level of control. In 1998 Loss and Divincenzo~\cite{loss1998} proposed the use of a controllable exchange interaction in semiconductor quantum dots to perform a two-qubit logic gate. In the same year Kane~\cite{kane1998} proposed how this could be achieved in donor based devices. Here, the wavefunction overlap between two electrons on neighbouring donor atoms placed ${\sim}20$~nm apart is controlled using an exchange gate between them. Harnessing this exchange interaction to perform a universal two-qubit quantum logic gate is the next step for donor based architectures. 

Three approaches exist for donor qubits: a controlled-phase (CZ) gate~\cite{Veldhorst:2015qv}; the controlled-rotation (C-ROT) gate~\cite{kalra2014} and a direct two-electron SWAP operation~\cite{nowack2011}. Whilst the first two protocols require the use of high frequency microwave fields for electron spin resonance~\cite{morello2010}, a direct two-electron SWAP necessitates the ability to turn on and off the exchange interaction between the electrons over orders of magnitude for high fidelity two-qubit operations. Notably, whilst the extent of a single donor wavefunction is well understood~\cite{koiller2002,Wellard2004,Wellard2005} modelling the exchange coupling between two donor electrons is more  complex~\cite{koiller2002,Wellard2004,Wellard2005,Saraiva2015,Gamble2015,Rahman2011} due to multi-valley interference effects~\cite{Cullis1970}. To this end, a critical challenge for donor based architectures is to know the distance required between the donors in order to turn the exchange interaction on and off with external gates~\cite{Saraiva2015,Gamble2015}. 

To date, two main methods for donor placement in silicon exist; ion implantation~\cite{jamieson2005} and atomic manipulation via scanning-tunnelling-microscopy (STM) hydrogen lithography~\cite{simmonsPRL2003}. Despite much success in accessing randomly placed donor spins, ion implantation has yet to demonstrate donor placement precision below $\sim6$~nm, whilst STM lithography has demonstrated donor placement down at the atomic scale~\cite{fuechsle2012}. 

In this paper we use STM lithography which allows both the precision placement of donor atoms for direct and independent spin-measurement of electrons near a readout structure and, most importantly, the control of the exchange interaction between them. We measure the anti-correlated spin states that arise due to the formation of two electron singlet-triplet states as a function of their wavefunction overlap, which is controlled by in-plane detuning gates. By observing the onset of these anti-correlated spin states as a function of detuning pulse voltage and time, we estimate the magnitude of tunnel coupling between the two donor qubits, and provide a roadmap toward coherent exchange gates for future devices. 

\section{Results}
\subsection{Independent spin readout of a 2P-1P system}

In the original Kane proposal an exchange gate between the donors was suggested to directly tune the exchange coupling between the qubits~\cite{kane1998}. Recent tight binding simulations have shown that it is difficult to tune the exchange energy in a 1P-1P donor configuration using such a gate~\cite{Wang2016}. Instead, it has been proposed that the exchange energy could be tuned over five orders of magnitude~\cite{Wang2016} by confining electrons in an asymmetric 2P-1P configuration and by utilising `tilt' control using two opposing detuning gates rather than a central J-gate, see Fig.~\ref{fig:readout}a. Motivated by these predictions with estimates for the required inter-donor separation, in this paper we demonstrate the ability to control exchange coupling in donor based qubits at the (1,1)-(2,0) charge region using a 2P-1P donor system.

The device, shown in Fig.~\ref{fig:readout}a, was patterned using STM hydrogen lithography. The qubits-L, and -R (left and right) are composed of $2$ donors and $1$ donor respectively, determined by examining the size of the lithographic patches (insets in Fig.~\ref{fig:readout}a) as well as their measured charging energies~\cite{buch2013,weber2014} (see Supplementary Figure 1 and Supplementary Figure 6). Three gates, \{G\textsubscript{L}, G\textsubscript{M},  G\textsubscript{R}\} control the electrostatic environment of the qubits which are  tunnel coupled to a larger readout structure made up of approximately 1000~P atoms which serves as a single-electron-transistor (SET) charge sensor. The SET quantum dot is operated with a source-drain bias of 2.5~mV, has a charging energy of ${\sim}6$~meV and is controlled predominantly via gate G\textsubscript{S}. All data in this paper was taken in a dilution refrigerator with a base temperature of $\sim$100~mK (electron temperature $\sim$200~mK). 

\begin{figure*}[t!]
\begin{center}
\includegraphics[width=1\textwidth]{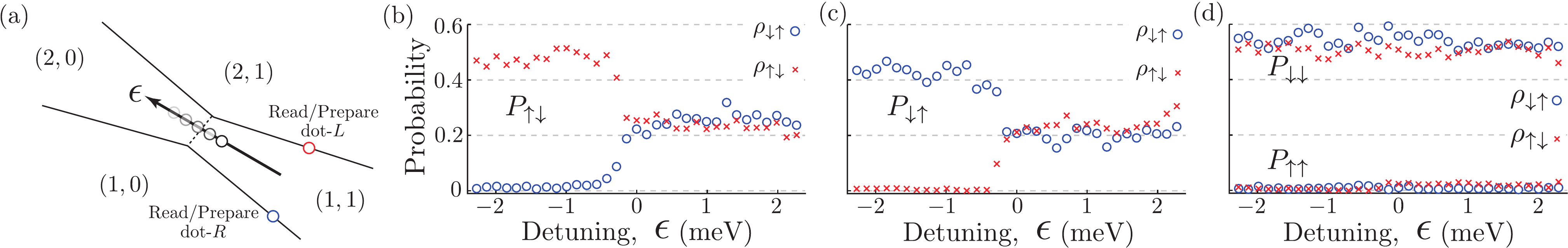}
\end{center}
\vspace{-0.5cm}
\caption{{\bf Controllable exchange interaction between precision placed donor-based spin qubits.} {\bf a}, We prepare a random spin on one qubit and deterministically spin-down on the other near the readout positions in the (1,1) charge region shown by the red and blue circles. After moving into the (1,1) region equidistant between the read positions for $1$~ms (start of arrow), a $50$~ms pulse is applied along the detuning axis shown by the black arrow to the positions marked by the black circles. Subsequent pulses are applied to perform spin readout on both qubits. {\bf b}, {\bf c} and {\bf d}, The probabilities of the joint two-spin outcomes from sequential spin-readout of L and R plotted against the detuning energy, $\epsilon$. For the initially prepared state $\rho_{\downarrow\uparrow}$ the blue circles show the outcome of two electron spin readout which is  performed at approximately $\epsilon{=}{-}7$~meV detuning in the (1,1) charge region where exchange is negligible (see Fig.~\ref{fig:readout}). The red crosses show the equivalent data set for an initially prepared state $\rho_{\uparrow\downarrow}$.}
\vspace{-0.5cm}
\label{fig:corr}
\end{figure*}

Figure~\ref{fig:readout}b shows the charge stability map of the 2P-1P device. Current peaks running diagonally correspond to charge transitions of the SET island. Two sets of breaks in the SET current peaks are observed with different slopes and correspond to electron transitions from either L or R to the SET island. An avoided crossing (triple-point) between these two transitions (white dotted line) indicates the region where electrons can tunnel between L and R, in this case at the (1,1)-(2,0) charge transition. Only one more charge transition corresponding to L is observed at lower gate voltages leading to the assignment of the charge regions. 

\label{sec:expt}

The direct measurement of anti-correlated electrons hinges upon the ability to independently measure their spin states~\cite{Watsone1602811}. To measure the spin of R we employ an energy-selective tunnelling technique~\cite{buch2013} where the electrochemical potential of the single-electron transition from the $1{\rightarrow}0$ charge state is split by the Zeeman energy in a static magnetic field $B_z$, see blue arrows in Fig.~\ref{fig:readout}c. Whether the electron is able to tunnel to the SET reservoir therefore depends on its spin state, i.e. the readout is a spin-dependent unloading mechanism from the qubit to SET. This readout technique is employed for the electron at R because the (1,1) region for this qubit borders the $1{\rightarrow}0$ charge states. 

For L we use a variant of this method, first reported Watson et al.~\cite{watson2015}. The charge transition for this qubit borders the $1{\rightarrow}2$ charge states, but because the chemical potential from the 1-electron spin-up and -down states to the 2-electron singlet state are also split by the Zeeman energy a similar readout method is allowed (red arrows in Fig.~\ref{fig:readout}c~\cite{Watsone1602811}). In this case, we utilise a spin-dependent loading mechanism from the SET to L. The combination of these two distinct readout techniques avoids the need to pulse over large voltages in order to reach the (1,1)-(2,0) charge transition. Both readout methods are equivalent and give rise to a current `blip' through the SET which is used to discriminate between spin-up and -down electrons. The average readout fidelity of spin-up and -down are estimated to be $96.2{\pm}1.1$\% and $97.6{\pm}2.1$\% for qubit-L and -R respectively (See Supplementary Figures 2, 3 and Table 1 for full analysis).

Importantly, the readout of each electron must be completely independent of the spin-state of the other. That is to say, the exchange energy at the detuning-position where readout is performed must be vanishingly small, such that no spin flips-flops occur during the readout window. This is demonstrated in Fig.~\ref{fig:readout}e and f. For these measurements we prepare one of two states, 
\begin{equation}
\rho_{\downarrow\uparrow} = \frac{\ket{{\downarrow\uparrow}}\bra{{\downarrow\uparrow}}{+}\ket{{\downarrow\downarrow}}\bra{{\downarrow\downarrow}}}{2} \text{  ,  } \rho_{\uparrow\downarrow} = \frac{\ket{{\uparrow\downarrow}}\bra{{\uparrow\downarrow}}{+}\ket{{\downarrow\downarrow}}\bra{{\downarrow\downarrow}}}{2}. 
\label{eq:rhoin}
\end{equation}
where $\ket{i,j}$ indicates the spin state $i$ and $j$ on qubit-L and -R respectively. Loading spin-down for one qubit is performed deterministically as a result of the spin readout protocol. Spin-up cannot be loaded deterministically, instead a random mixture of spin up and down is loaded by plunging the qubit far below the SET fermi-level. After initialisation we pulse inside the (1,1) charge region midway between the two readout positions (green square in Fig.~\ref{fig:readout}b) and wait for up to 10~s for the randomly loaded electron spin to decay to spin down. Sequential spin-readout of L and then R is performed, in that order, to minimise the effect of the shorter $T_1$ of qubit-L. The spin-up fractions show relaxation of the qubit initially loaded with random spin, with $T_1$ times measured to be $2.9\pm0.5$~s and $9.3\pm2.4$~s for electrons on L and R respectively at $B_z{=}$2.5T. Importantly, the electron initially loaded as spin-down shows no significant spin-up fraction during this time, demonstrating that at these readout positions there is no significant spin-spin interaction over $\sim10$~s. 

\subsection{Controllable exchange of precision placed donors}

The realisation of a two-qubit logic gate hinges on the ability to controllably turn on and off an interaction between quantum particles. We show this here by pulsing toward the (1,1)-(2,0) charge transition where an exchange interaction between the two electrons arises as a consequence of the Pauli-exclusion principle~\cite{dirac1926}. The Hamiltonian is given by $H_{\textrm{ex}}{=}J\bold{S_{L}}{\cdot}\bold{S_{R}}$, where $\bold{S_{L}}$ and $\bold{S_{R}}$ are the left and right electron spin vectors and $J$ is the strength of the exchange interaction~\cite{petta2005}. The magnitude of $J$ is given by the energy difference between the symmetric and anti-symmetric two-electron states $\ket{T^0}{=}\left(\ket{{\uparrow\downarrow}}{+}\ket{{\downarrow\uparrow}}\right)/\sqrt{2}$ and $\ket{S}{=}\left(\ket{{\uparrow\downarrow}}{-}\ket{{\downarrow\uparrow}}\right)/\sqrt{2}$ respectively. Similar to gate defined quantum dots, it has been shown that the exchange between donors can also be parameterised in terms of the tunnel coupling and detuning between the (1,1) and (2,0) charge states~\cite{House:2015rz},
\begin{equation}
J(\epsilon)=\frac{\epsilon}{2} + \sqrt{t_\textrm{c}^2 + \left(\frac{\epsilon}{2}\right)^2}, 
\label{eq:J}
\end{equation} 
where $\epsilon$ is the detuning and $t_\textrm{c}$ is the tunnel coupling (such that $J(0){=}t_\textrm{c}$). The detuning axis $\epsilon$ is applied along $V_{\textrm{GL}}{=}{-}0.9V_{\textrm{GR}}$ (along the SET Coulomb blockade) and is shown by the white arrow in Fig.~\ref{fig:readout}a. It effects a tilting from the (1,1) toward the (2,0) charge state, shown schematically in Fig.~1d. The detuning energy, $\epsilon$, is related to the applied gate voltage $V_{\textrm{GL}}$ via the lever arm $\alpha_{\epsilon}{=}0.071$~eV/V, such that $\epsilon{=}\alpha_{\epsilon}V_{\textrm{GL}}$.

\begin{figure}
\begin{center}
\includegraphics[width=1\columnwidth]{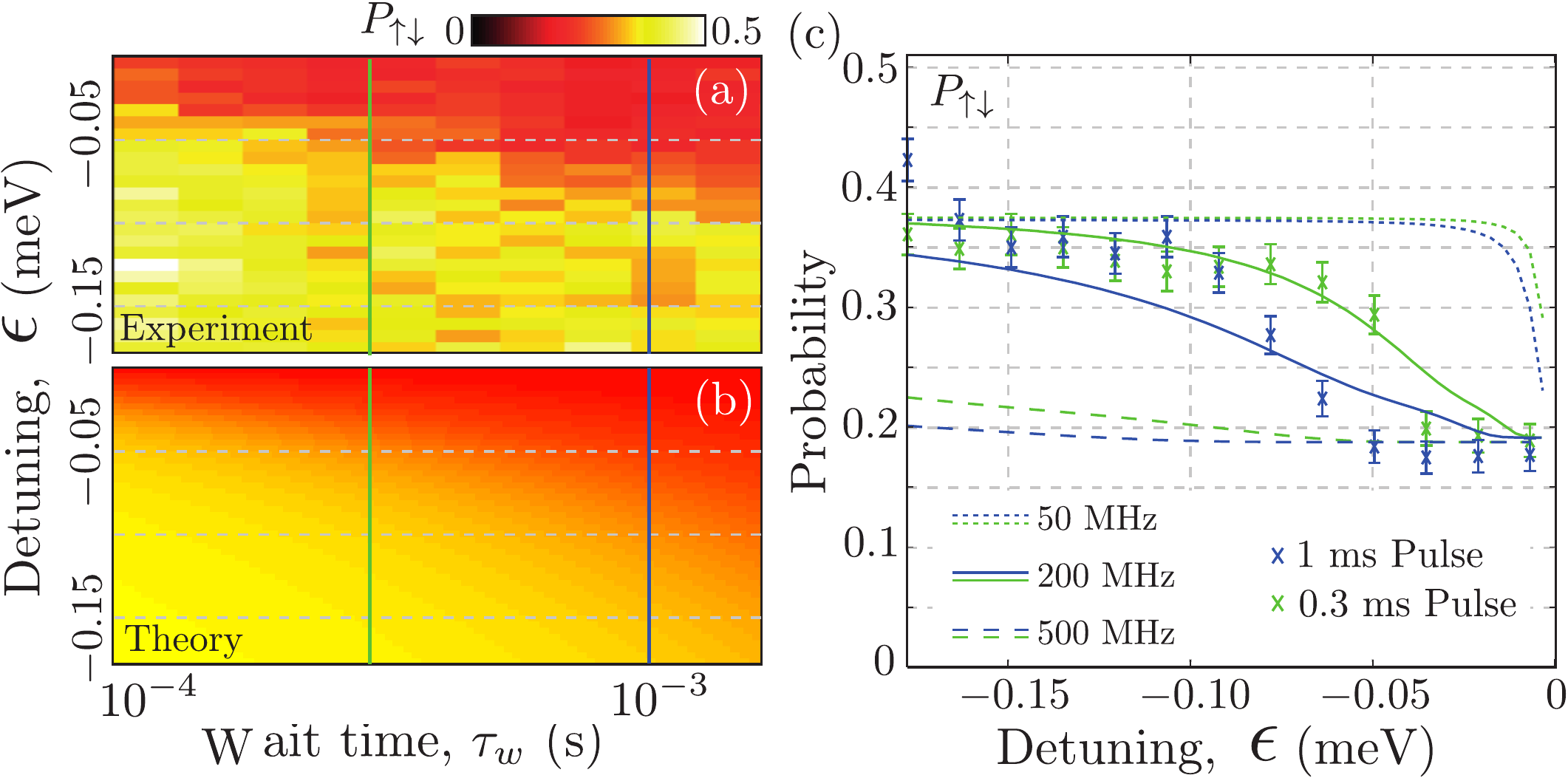}
\end{center}
\vspace{-0.5cm}
\caption{{\bf Experimental estimation of tunnel coupling.} Starting with $\rho_{\uparrow\downarrow}$, the measured probability $P_{\uparrow\downarrow}$ ({\bf a}) and the theoretical prediction ({\bf b}) as a function of pulse wait time and detuning position. For the model we have used a value of $t_\textrm{c}{=}200$~MHz. {\bf c}, Solid green and blue curves show theoretical predictions for wait times of 0.3 and 1~ms respectively (corresponding cuts shown in the lower plot of {\bf a}). Blue and green crosses show measurements for these wait times. The dashed and dotted lines show the theoretical predictions for tunnel coupling values of $t_\textrm{c}{=}$500~MHz and 50~MHz respectively.}
\vspace{-0.5cm}
\label{fig:corrtheory}
\end{figure}

We start by initialising either state from Eq.~\ref{eq:rhoin} by loading one qubit randomly and deterministically down on the other, and subsequently apply a $50$~ms pulse along the axis $\epsilon$ to control the strength of the exchange interaction~\cite{petta2005}, shown by the open black circles in Fig.~\ref{fig:corr}a. This time is long enough to allow for a significant exchange interaction, but much shorter than any electron spin relaxation such that readout is not hindered. Upon pulsing back into the (1,1) charge region we perform independent spin readout of L and then R. In addition to the single spin outcomes for each qubit we also determine the joint probabilities $P_{ij}$ for ${ij{\in}\{\uparrow\uparrow,\uparrow\downarrow,\downarrow\uparrow,\uparrow\uparrow\}}$, as shown in Fig.~\ref{fig:corr}b-d. 

\begin{figure*}
\begin{center}
\includegraphics[width=1\textwidth]{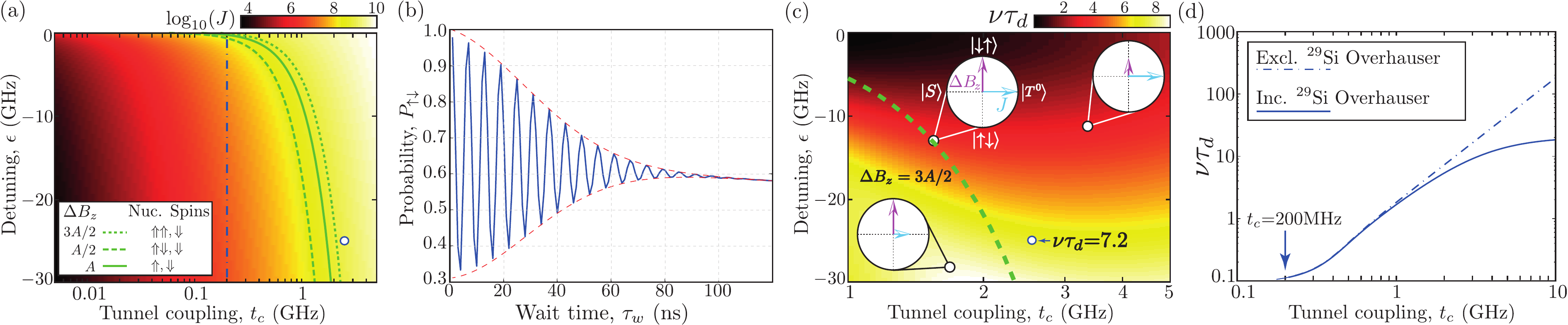}
\end{center}
\vspace{-0.5cm}
\caption{{\bf Theoretical predictions for the observation of coherent exchange oscillations.} {\bf a}, The value of exchange energy, $J$, as a function of tunnel coupling, $t_\textrm{c}$ and detuning, $\epsilon$. The boundary separating the two-electron product states with singlet-triplet states occurs where the difference in magnetic field between the two qubits, $\Delta B_z$ is equal to the exchange energy. For donor based systems $\Delta B_z$ is dominated by the donor hyperfine strength, and is equal to $A$ for a 1P-1P system (solid green line), and can take the two values $A/2$ or $3A/2$ for a 2P-1P system (dashed and dotted green lines respectively) dependent on the nuclear spin orientation (examples shown in inset). We assume the bulk 1P value for the hyperfine, $A{=}117.53$~MHz. The dashed blue line indicates the values of $J$ accessible for the current device with $t_\textrm{c}{\sim}200$~MHz. {\bf b}, Theoretical prediction of coherent exchange oscillations for a 2P-1P device in natural silicon with tunnel coupling $t_\textrm{c}{=}2.5$~GHz. The two electron state is initialised as $\ket{{\uparrow\downarrow}}$ at a point where the exchange energy is negligible, and subsequently a non-adiabatic detuning pulse is applied to $\epsilon{=}-25$~GHz (circle marker in {\bf a}). We have assumed voltage noise equivalent to 850~MHz along the detuning axis, $\epsilon$  (obtained from measurements) as well as a single electron $T^{*}_2{=}55$~ns measured in previous works~\cite{morello2010}. From this result an oscillation frequency $\nu$ and dephasing time $\tau_\textrm{d}$ are extracted. {\bf c}, The product of oscillation frequency, $\nu$ and dephasing time, $\tau_\textrm{d}$ as a function of tunnel coupling and detuning. The green dashed line represents the boundary between product and singlet-triplet eigenstates of the two-electron system. The Bloch sphere cross sections indicate the relative magnitudes of $\Delta B_z$ (purple) and $J$ (blue) in different regions. {\bf d}, Theoretical prediction of $\nu\tau_\textrm{d}$ along the line $\Delta B_z{=}J$ as a function of tunnel coupling for a 2P-1P double quantum dot. Solid (dashed) line shows results including (excluding) the $^{29}$Si Overhauser field.}
\vspace{-0.5cm}
\label{fig:Model}
\end{figure*}

In the case where $\rho_{\uparrow\downarrow}$ is initialised, after pulsing to $\epsilon{\ll}0$ we observe $P_{\uparrow\downarrow}{\sim}0.5$ and $P_{\downarrow\uparrow}{\sim}0$, indicating no spin flip-flops have occurred during the 50~ms pulse duration. However, anti-correlated spins can be seen in Fig.~\ref{fig:corr}(b,c) as we pulse closer to the (1,1)-(2,0) charge transition, at $\epsilon{=}0$ where both $P_{\uparrow\downarrow}$ and $P_{\downarrow\uparrow}{\rightarrow}0.25$. Furthermore, we see that both $P_{\uparrow\uparrow}$ and $P_{\downarrow\downarrow}$ remain constant at approximately 0 and 0.5 respectively as they represent populations of the triplet states $\ket{{\uparrow\uparrow}}$ and $\ket{{\downarrow\downarrow}}$ and are not subject to the exchange interaction. Statistical analysis (see Supplementary Figure 4) of these results indicates a correlation coefficient of $\phi{=}{-}0.243{\pm}0.028$ with a p-value ${\ll}0.01$ for $0{<}\epsilon{<}2.4$~meV, demonstrating the presence of statistically significant spin anti-correlations in this region. 

\subsection{Estimate of inter-donor exchange coupling}

To ascertain the value of the inter-dot tunnel coupling, $t_\textrm{c}$, we repeat the same pulsing scheme as above whilst modifying the detuning pulse duration from 0.1-2~ms and compare our results to a spin-level theoretical model, see Fig.~\ref{fig:corrtheory}. The quantum mechanical behaviour of a donor based two-qubit system is described by the following terms in the Hamiltonian:
\begin{equation}
\begin{split}
				H_{\textrm{ze}}&=\gamma_\textrm{e} \bold{B}\cdot(\bold{S_{L}} + \bold{S_{R}}), \\
				H_{\textrm{zn}}&=\gamma_\textrm{n} \bold{B}\cdot\left(\bold{I_{n_{L1}}} + \bold{I_{n_{L2}}} + \bold{I_{n_R}}\right), \\
				H_{\textrm{hf}}&=A_\textrm{L}\bold{S_{L}}\cdot\left(\bold{I_{n_{L1}}} + \bold{I_{n_{L2}}}\right) + A_\textrm{R}\bold{S_{R}}\cdot\bold{I_{n_R}}, \\
				H_{\textrm{ex}}&=J\bold{S_{L}}\cdot\bold{S_{R}},
\label{eq:SpinHam}
\end{split}
\end{equation}
where $H_{\textrm{ze}}$ and $H_{\textrm{zn}}$ are the electron and nuclear Zeeman energies, with $\gamma_\textrm{e}{=}28.024$ GHz/T and $\gamma_\textrm{n}{=}17.235$ MHz/T gyromagnetic ratios respectively~\cite{weast1988crc}. The hyperfine term, $H_{\textrm{hf}}$ is separated into two components as it has been predicted that the hyperfine constants will be different for varying donor cluster configurations~\cite{buch2013,Rahman2007}. Here for simplicity we assume the bulk-like value of $A_\textrm{L}{=}A_\textrm{R}{=}A{=}117.53$ MHz~\cite{weast1988crc} and define the static field to be $\bold{B}{=}(0,0,|B_z|)$. We numerically calculate the time evolution of the density matrix via a 4th order Runge-Kutta method with the inclusion of relevant decoherence channels (see Supplementary Figure 5). 

For the theoretical data shown in the lower panel of Fig.~\ref{fig:corrtheory}a we prepare the initial state $\rho_{\uparrow\downarrow}$ from Eq.~\ref{eq:rhoin} and simulate non-adiabatic pulses to detuning positions for varying pulse durations, $\tau_\textrm{w}$. For this simulation we use a tunnel coupling, $t_\textrm{c}{=}200$~MHz, assume dynamic P nuclear spins as well as a single-spin dephasing time of $T_2^*{=}55$~ns due to the constantly fluctuating Overhauser field of the $^{29}$Si nuclear spins~\cite{pla2012}. The equivalent measured data set is shown in the upper panel of Fig.~\ref{fig:corrtheory}a with cuts at 0.3 and 1~ms shown in Fig.~\ref{fig:corrtheory}b and compared with the theoretical predictions for $t_\textrm{c}{=}50$, 200 and 500~MHz. From these results we can estimate the tunnel coupling within an order of magnitude accuracy to be $t_\textrm{c}{\sim}200$~MHz for this device. Following Eq.~\ref{eq:J}, this result places an equivalent bound on the achievable exchange energy $J{<}$200~MHz inside the (1,1) charge region (see Fig.~\ref{fig:Model}a).

In Ref.~\cite{Wang2016} the authors investigated multiple different 2P intradot configurations, and found that disorder at the lattice-site level had little effect on the final exchange energy. They showed that the exchange energy for a 2P-1P system with a 15nm separation was tunable over five orders of magnitude for electric field strengths $-2{<}|\bold{E}|{<}2$~MV/m at the donor sites. The voltage applied to the in-plane gates in our device amount to a potential difference of $\sim$100~mV between $V_{\textrm{GL}}$ and $V_{\textrm{GR}}$ at the (1,1)-(2,0) inter-dot transition. From an electrostatic model of our device we estimate $|\bold{E}|{=}0.49{\pm}0.10$~MV/m at the donor sites. Our estimate of $t_\textrm{c}{=}0.2$~GHz (which is equal to $J$ at $\epsilon{=}0$) is within an order of magnitude of the theoretical prediction for $J$ in a 2P-1P system given this $|\bold{E}|$~\cite{Wang2016}.

\subsection{Requirements for coherent control of exchange}

In this final section we investigate the potential for achieving coherent exchange between two electrons confined to donors in natural silicon. Based on Eq.~\ref{eq:J} the plot in Fig.~\ref{fig:Model}a shows the obtainable exchange energies, $J$, as a function of detuning and tunnel coupling, where the vertical  blue dashed line shows $t_\textrm{c}{=}0.2$~GHz for our device. Importantly, the boundary where the difference in magnetic field at the two qubit sites $\Delta B_z$ is equal to the exchange $J$, separates the two-electron product eigenstates and singlet-triplet eigenstates. For donor qubits $\Delta B_z$ is dominated by the phosphorus nuclear-spin hyperfine, $A$. The exact value of $\Delta B_z$ varies depending on the number of donors at each qubit site and their nuclear spin orientations: For a 2P-1P device with random nuclear spin configurations $\Delta B_z$ fluctuates between $3A/2$ or $A/2$ with a 1:3 ratio.

It can be seen from Fig.~\ref{fig:Model}a that for the device studied here there exists only a small range in detuning (approximately $10~\mu$V in gate voltage) over which one could implement coherent exchange oscillations inside the (1,1) charge region (negative $\epsilon$). When one takes into account any voltage noise on gates (which influences $\epsilon$ and ultimately $J$) this makes the operation of coherent oscillations challenging~\cite{Hu2006}. Indeed, in this particular device we measured gate RMS voltage noise of $50~\mu$V from shot to shot, equivalent to detuning noise of $\delta\epsilon{=}$850~MHz, indicating that pulsing repeatedly to the same exchange energy would not be possible for $\epsilon{<}0$. For the same reasons, charge noise also destroys coherence when adopting the approach to pulse $\epsilon{>}0$. We carried out experiments with pulses down to 10~ns for both $\epsilon{<}0$ and $\epsilon{>}0$ but were unable to observe coherent exchange phenomenon. 

Figure~\ref{fig:Model}b shows the predicted number of exchange oscillations (${\sim}15$) that would be observed in a device with $t_\textrm{c}{=}$2.5~GHz after pulsing to a detuning $\epsilon{=}{-}25$~GHz (circle marker in Fig.~\ref{fig:Model}a). Conversely, using the same model we estimate that a noise floor of ${<}6~\upmu$V (${\sim}100$~MHz in detuning or ${\sim}50$~mK, much lower than the electron temperature) would be required to observe the signature of coherent oscillations in the present device. Note that in addition to the realistic detuning noise we have also included the effect of a constantly fluctuating $^{29}$Si Overhauser field expected in natural silicon~\cite{morello2010} as well as randomized P donor nuclear spins of the donor atoms themselves.

From these simulations we can extract the frequency of oscillations, $\nu$ as well as the dephasing time $\tau_\textrm{d}$, allowing us to determine the figure of merit $\nu\tau_\textrm{d}$ as a function of tunnel coupling and detuning pulse position, see Fig.~\ref{fig:Model}c. Interestingly, the product $\nu\tau_\textrm{d}$ only becomes significant beyond the boundary $\Delta B_z{=}3A/2$ for values of $t_\textrm{c}{>}2$~GHz, providing a lower bound on the required tunnel coupling for coherent control. Figure~\ref{fig:Model}d gives $\nu\tau_\textrm{d}$ as a function of tunnel coupling for a detuning pulse to the boundary $\Delta B_z{=}3A/2$. These results indicate that at high tunnel coupling, the observation of exchange oscillations will ultimately be hindered by the presence of the fluctuating $^{29}$Si Overhauser field. In the case where qubits exist in a spin vacuum, as in $^{28}$Si, only charge noise is relevant and $\nu\tau_\textrm{d}$ can be seen to increase monotonically as a function of tunnel coupling (dashed line in Fig.~\ref{fig:Model}d). 

\section{Discussion}

In summary, we have demonstrated a controllable exchange interaction resulting in two-electron spin anti-correlations on precision placed 2P-1P donors qubits in Si using in-plane `detuning' gates. The results are consistent with the exchange interaction behaviour expected at the (1,1)-(2,0) charge transition and represent the first direct measurement of correlated electron spins in donor based devices. Whilst the small tunnel coupling (0.2~GHz) in the present device prohibited measurement of coherent oscillations, we show our results agree with recent studies~\cite{Saraiva2015} in which much smaller distances than previously predicted are required to achieve a sufficiently large exchange coupling for coherent control. Furthermore, while detuning noise presents a problem for devices with a small tunnel coupling, we theoretically predict that for larger tunnel couplings of $t_\textrm{c}{>}2$~GHz it can be overcome. Theoretical work on coupled donor systems suggest a separation of 13-14nm between a 1P-2P system will be required to achieve this magnitude of exchange coupling~\cite{Wang2016}. Importantly, there is no reason to believe that this small change in donor site separation will lead to a significant reduction in electrical controllability based on previous experimental works~\cite{doi:10.1021/nl3012903,weber2014,House:2015rz,Watsone1602811}. This benchmark for a larger interaction strength between neighbouring donor-based qubits provides the focus for future experiments. 

With the atomic precision placement of donors using STM lithography it will be possible to further optimise the inter-donor distance to control the coherent coupling between two donors qubits with order-of-magnitude accuracy~\cite{koiller2002b,koiller2002b}. Whilst extensive studies have been conducted for deterministic single P donor incorporation~\cite{bussmann2014imaging} similar studies will need to be developed to determine the optimal lithographic patch for deterministic 2P incorporation. Crucially, recent theory predicts that the 1P-2P configuration we present in this paper both increases the tunability of the tunnel coupling and at the same time suppresses the `exchange fluctuations' known for two single donors, and may therefore be less sensitive to the exact atomistic donor positions than two coupled single donors~\cite{Wang2016}. Furthermore, our ability to directly place the donor with ${<}1$~nm accuracy along with the reproducible demonstration of high fidelity single-shot spin-readout in multiple devices~\cite{Watsone1602811}, bode well for the future scalability of donor qubit quantum computers. 

\section{Methods}
\vspace{-0.25cm}
\subsection{Device fabrication}
\vspace{-0.5cm}
\noindent The device, shown in Fig.~\ref{fig:readout} was fabricated using scanning-tunnelling-microscopy hydrogen lithography to selectively remove hydrogen from a passivated Si(100) $2{\times}1$ reconstructed surface. The lithographic mask is subsequently dosed with PH$_3$ and annealed (320$^\circ$C) to incorporate P atoms into the silicon substrate~\cite{fuhrer2009} with $\sim$1/4~ML density (~$2\times 10 ^{14}$cm$^{-2}$) allowing for quasi-metallic conduction in all electrodes~\cite{weber2012b}. 
\vspace{-0.5cm}
\subsection{Measurement setup}
\vspace{-0.5cm}
\noindent For all electrical measurements, the device was mounted on a high-frequency printed circuit board within a copper enclosure, thermally anchored to the cold finger of a dilution refrigerator with a base temperature of 50~mK. Voltage pulses were applied to gates G\textsubscript{L} and G\textsubscript{R} by an arbitrary waveform generator (Agilent 81180A), connected via a bias tee to the gate along with a constant-voltage source. The SET current, $I_\textrm{SET}$, was amplified and converted into a voltage signal at room temperature, low-pass filtered to 1 kHz bandwidth, and acquired with a fast digitizing oscilloscope.

\subsection{Data availability}
The data that support the findings of this study are available from the corresponding author upon reasonable request.

\section{Author Contributions}
\noindent M.A.B., S.K.G., J.G.K. and W.B. fabricated the device. M.A.B., S.K.G., M.H. and S.J.H. obtained all measurements. Calculation of donor charging energies was performed by S.J.H. and S.K.G. The data analysed by M.A.B., S.K.G. and M.H. and was discussed critically with all authors. Theoretical modelling of coherent exchange coupling was carried out by M.A.B. and D.K. with input from S.K.G. and C.D.H. Spin-readout analysis was performed by M.A.B., D.K. and T.F.W. The manuscript was written by M.A.B., S.K.G and M.Y.S. with input from all other authors. M.Y.S. supervised the project. 

\vspace{1cm}
\section{Competing Financial Interests}
\noindent All authors declare no competing financial interests.

\section{acknowledgements}
\noindent We thank Sven Rogge for enlightening discussions. This research was conducted by the Australian Research Council Centre of Excellence for Quantum Computation and Communication Technology (project no. CE110001027) and the US National Security Agency and US Army Research Office (contract no. W911NF-08-1-0527). M.Y.S. acknowledges an ARC Laureate Fellowship.


\setcounter{figure}{0}
\makeatletter 
\renewcommand{\thefigure}{S\@arabic\c@figure}
\makeatother

\begin{widetext}
\vspace{10cm}

\section{\large{Supplemetary information}}

\begin{figure}
\begin{center}
\includegraphics[width=0.9\textwidth]{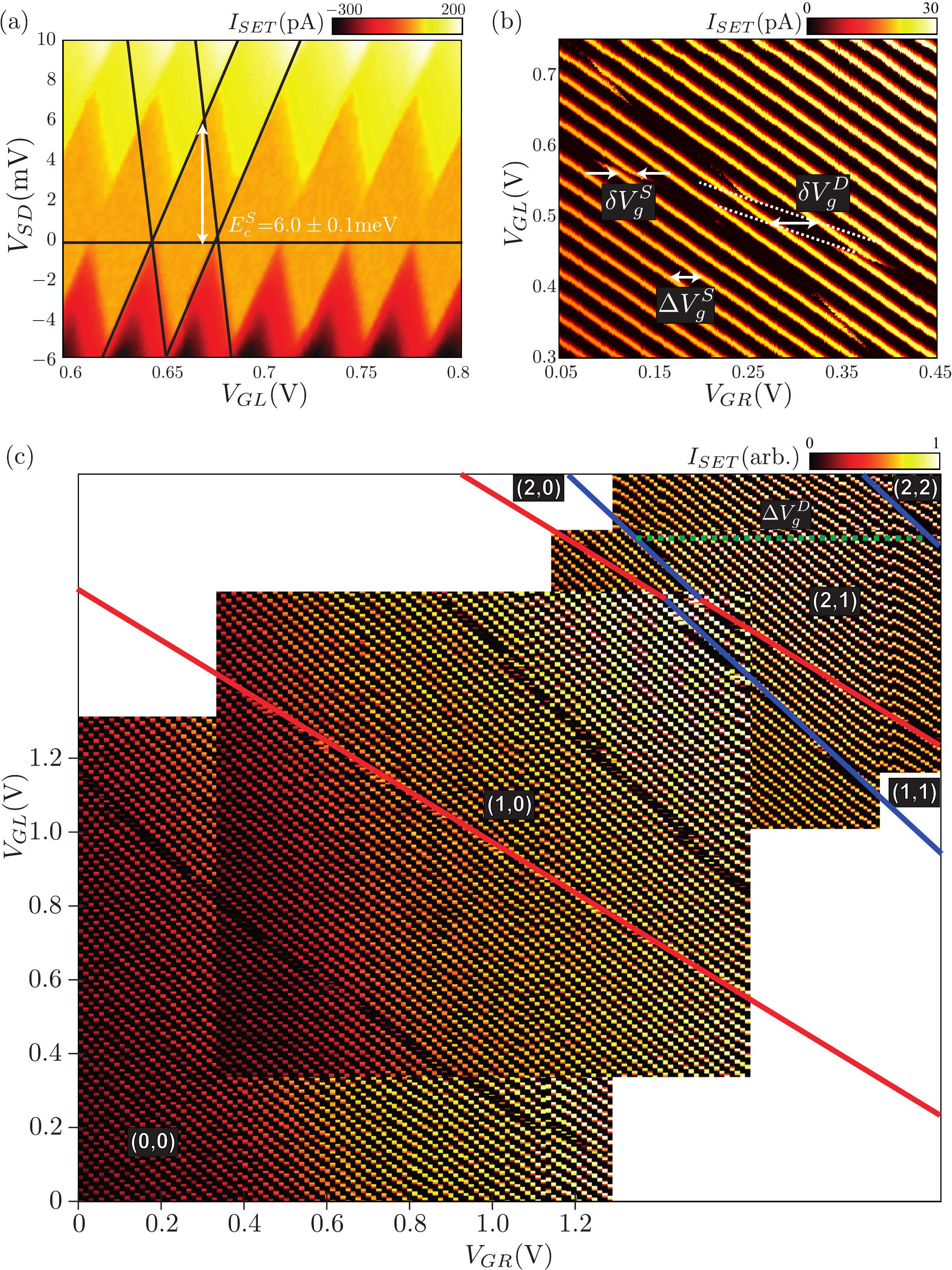}
\end{center}
\caption{{\bf Definition of the parameters used in calculating the charging energy for the two qubits.} {\bf a}, The Coulomb diamonds used to extract the charging energy of the SET, $E_c^{S}$. {\bf b}, A charge stability map ($V_{GL}$ vs. $V_{GR}$) showing the definition of the voltage parameters, $\Delta V_g^S$, $\delta V_g^S$ and $\delta V_g^S$. {\bf c}, A composite charge stability map showing all the observed charge transitions in our device as a function of $V_{GL}$ and $V_{GR}$. The four separate maps vary in the middle gate voltage, $V_{GM}$, from $0.1{-}0.7$~V. Red and blue lines show charge transitions of qubits $L$ and $R$ respectively. Other observed transitions are attributed to charge traps in the vicinity of the SET and are not relevant to the experiment. The definition of $\Delta V_g^D$ is shown by the green dashed line. The number of SET charge transitions, $n^S$ are counted along this line.}
\label{fig:def}
\end{figure}

\section{Charging energy calculation}
\label{sec:energies}
The charging energy, $E_c^a$ of a quantum dot (QD) $a$ can be calculated by knowing the charging energy of another QD, $b$ which is capacitively coupled to $a$~\cite{hile2015}. They are related through their mutual charging energy, $E_m$, which is given by,
\begin{equation}
E_m = \alpha_g^i \delta V_g^i,
\end{equation}
where $\alpha_g^i$ is the lever arm from gate $g$ to QD $i$ and $\delta V_g^i$ is the voltage shift of the potential of QD $i$ due to QD $j$. This value must be the same for both QDs $a$ and $b$. Therefore, we can write,
\begin{equation}
\alpha_g^a \delta V_g^a = \alpha_g^b \delta V_g^b.
\end{equation}
In the same manner, $E_c^i$ is given by,
\begin{equation}
E_c^i = \alpha_g^i \Delta V_g^i,
\end{equation}
where now, $\Delta V_g^i$ is the voltage difference between two charge transitions in the space of gate $g$. We can eliminate the (usually) unknown lever arms by combing equations Eq.(2) and Eq.(3),
\begin{equation}
\frac{E_c^a}{\Delta V_g^a} \delta V_g^a = \frac{E_c^b}{\Delta V_g^b} \delta V_g^b.
\end{equation}

We now consider the possibility that $E_c^a \neq E_c^b$ such that multiple charging events can occur in the voltage range $\Delta V_g^i$. In this case, the measured $\Delta V_g^i$ is actually the sum of the true $\Delta V_g^i$, which we denote $\Delta \hat{V}_g^i$ and the number of charging events, $n^j$ of the other QD,
\begin{equation}
\Delta V_g^i = \Delta \hat{V}_g^i + n^j \delta V_g^i,
\end{equation}
Since we require $\Delta \hat{V}_g^i$ in Eq.(4), we must substitute Eq.(5) in such that it now reads,
\begin{equation}
\frac{E_c^a}{E_c^b}  = \frac{\delta V_g^b(\Delta V_g^a - n^b \delta V_g^a)}{\delta V_g^a(\Delta V_g^b - n^a \delta V_g^b)}.
\end{equation}

\begin{figure}[b!]
\begin{center}
\includegraphics[width=0.8\textwidth]{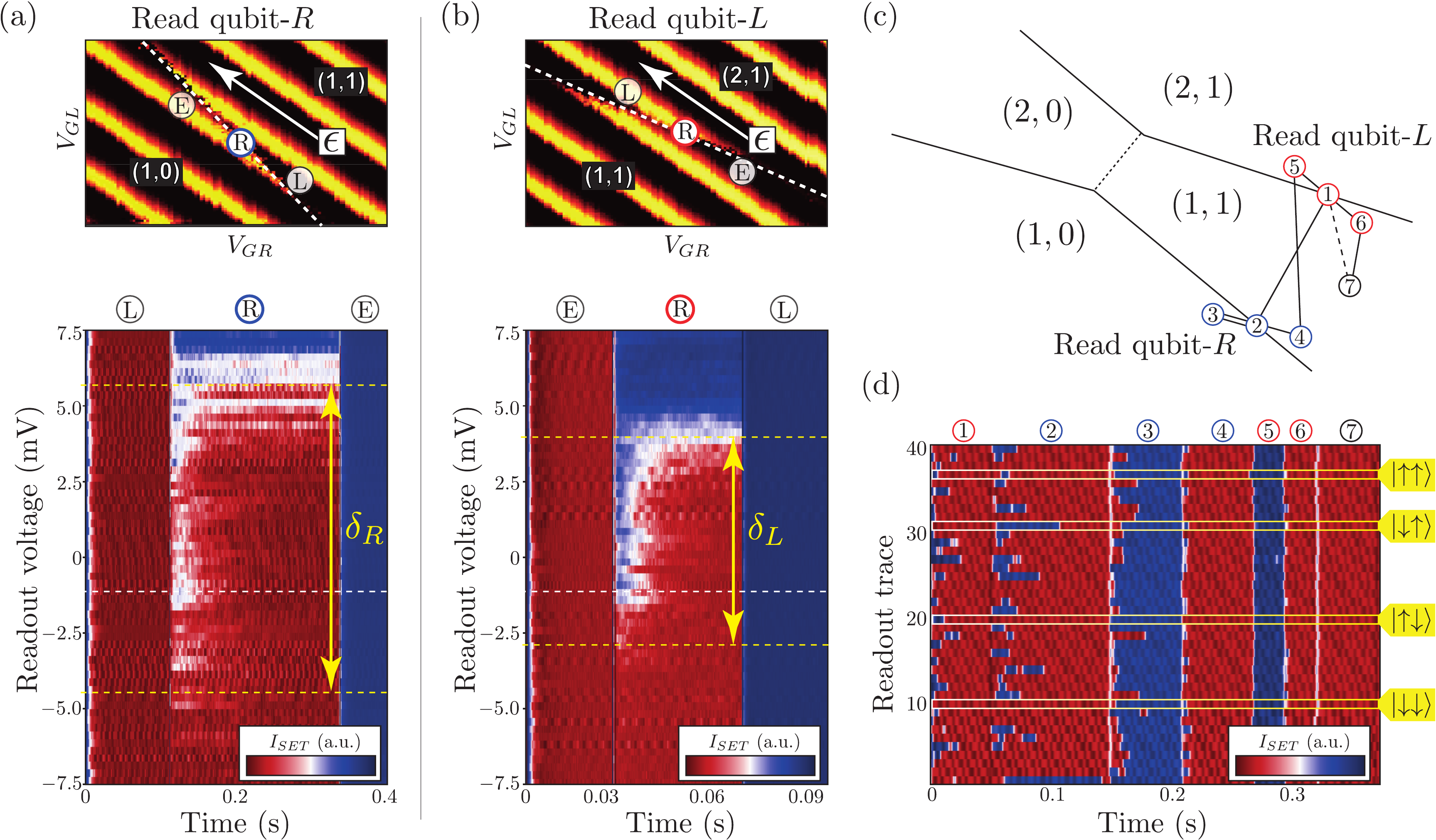}
\end{center}
\caption{{\bf Sequential spin readout of two donors}. {\bf a}, (upper) The three level pulse scheme for electron spin readout of qubit $R$ depicted on the charge stability map at the $1{\rightarrow}0$ charge transition (white dashed line). The approximate positions of the load, read and empty phases (R, L, E) of the three level pulse are shown by the circle markers. Spin readout of the electron at this donor relies on a spin-dependent \textit{unloading} mechanism from the qubit to the SET reservoir at the position marked `R' in the diagram~\cite{elzerman2004}. The read voltage is stepped along the SET Coulomb blockade peak, shown by the white arrow, and is equivalent to $\epsilon$ axis described in the main text. (lower) The average of 200 single-shot SET current traces, $I_{SET}$, as a function of the read voltage along $\epsilon$. The voltage at which spin readout is performed during the experiments is shown by the white dashed line. The range over which spin-up electrons can selectively tunnel off of the dot, $\delta_R$, is shown by the yellow arrow. All measurements were performed at $B_z{=}2.5$~T. {\bf b}, A similar readout method is employed for qubit $L$. Here, a spin-dependent \textit{loading} mechanism from the $1{\rightarrow}2$ charge state is utilised. {\bf c}, Schematic of the pulsing sequence used to sequentially readout $L$ and $R$ in that order, as well as initialise both qubits with random spin states. The order of spin readout is chosen to minimise the effects of spin relaxation since $L$ has the shorter $T_1$ time. To initialise spin-down deterministically on either qubit, we skip phases $3,4$ for $R$ and/or $5,6$ for $L$. {\bf d}, An example of 40 single shot traces for the sequential spin readout and initialisation of random spins on $L$ and $R$ for the sequence shown in {\bf c}. The two read phases occur at the beginning of each trace. A short `blip' in the SET current indicates the tunnelling of an electron during the read phases (1 and 2) in approximately $50\%$ of the traces. This occurs due to the presence of a spin-up electron on the dot. Example traces for the outcomes $\{\ket{{\uparrow\uparrow}},\ket{{\downarrow\uparrow}},\ket{{\uparrow\downarrow}},\ket{{\downarrow\downarrow}}\}$ are shown.}
\label{fig:readout}
\end{figure}

This is the general form of the equation which relates the two charging energies between QDs $a$ and $b$. We now look at the specific case where $\Delta V_g^a \gg \Delta V_g^b$. This is the case most commonly seen when we have a large QD used as a charge sensor to measure smaller QDs or single donors (which will necessarily have larger charging energies than the charge sensor QD). For clarity, we switch to the S/D (SET/donor) terminology in place of $a (= D)$ and $b (= S)$. If the condition $\Delta V_g^D \gg \Delta V_g^S$ holds then there will be multiple charging events of the SET ($n^S \neq 0$) and exactly zero for the donor ($n^D = 0$) within the voltage ranges $\Delta V_g^D$ and $\Delta V_g^S$, respectively. As a result, we can simplify Eq.(6) to,
\begin{equation}
E_c^{D} = \frac{E_c^{S}\delta V_g^{S}}{\Delta V_g^{S}} \Big(\frac{\Delta V_g^{D}}{\delta V_g^{D}} - n^{S}\Big).
\end{equation}
Figure~\ref{fig:def} shows the measurement of all required parameters $\{E_c^S,\delta V_g^S, \delta V_g^D, \Delta V_g^S, \Delta V_g^D\}$, where we have chosen to measure along the right gate, $G_R$, i.e. $g{\rightarrow}R$. Using Eq.(7) for the $1{\rightarrow}2$ electron transitions for both qubits, we find charging energies of $65{\pm}8$ and $43{\pm}5$ meV and for $L$ (2P) and $R$ (1P), respectively. These values are consistent with theoretical~\cite{saraiva2015,weber2014} and previously measured~\cite{fuechsle2012,weber2014} charging energies for 2P and 1P donor qubits respectively.

\begin{figure}[t!]
\begin{center}
\includegraphics[width=0.65\textwidth]{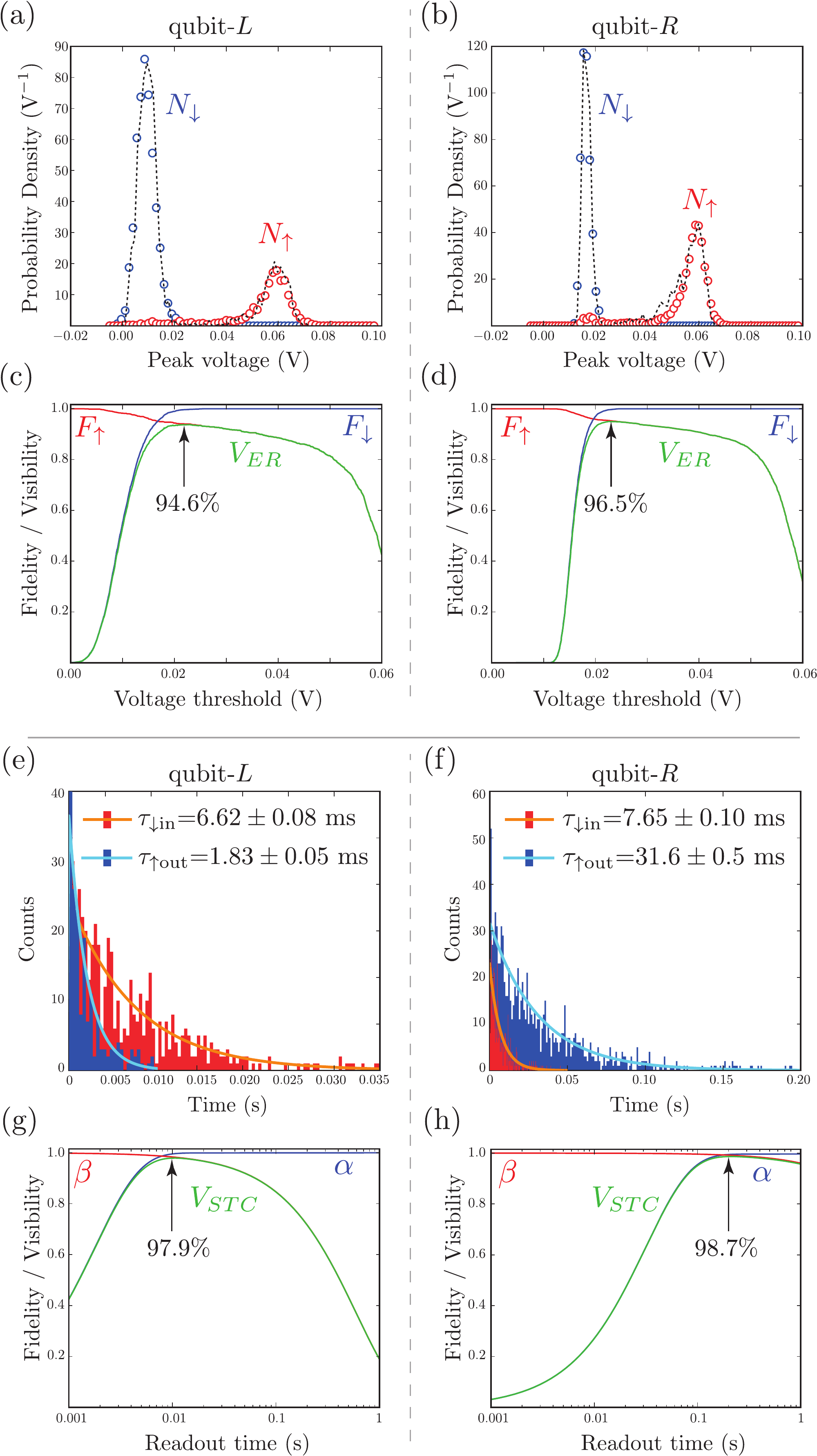}
\end{center}
\caption{{\bf Readout fidelity analysis.} {\bf a-d}, Calculation of electrical readout fidelity, $V_{ER}$. A simulation of $10,000$ SET traces during readout with $50\%$ containing `blips' and with the same signal-to-noise ratio as measured in the experiment, allows for an optimisation of the readout voltage threshold, $V_t$ being where $V_{ER}$ is maximised. Blue and red circles in {\bf a,b} show the simulated spin-up and -down traces respectively, whereas the dashed black line shows the experimental distribution of maximum voltage during the readout phases. The fidelities $F_{\uparrow}$ and $F_{\downarrow}$ in {\bf c,d} are calculated using Eq.~\ref{eq:rofids} as a function of the readout threshold $V_t$. {\bf e-h}, Calculation of the spin-to-charge conversion fidelity, $V_{STC}$. The relevant tunnel times from $L$ and $R$ are measured experimentally from $2,000$ SET readout traces shown in {\bf e,f}. A rate equation model developed in Refs.~\cite{buch2013,watsonthesis} is used to estimate the optimum readout time, $\Delta t$, which optimises the successful assignment of a spin-up or -down, $\alpha$ and $\beta$ respectively. From these individual fidelities the maximum spin-to-charge conversion fidelity, $V_{STC}$ can be estimated for both qubits, shown in {\bf g,h}.}
\label{fig:fidelity}
\end{figure}

\section{Sequential spin readout of two donor qubits}
The top panels of Fig.~\ref{fig:readout}a and b show close-ups of the current through the SET charge sensor in the region where spin readout is performed on $R$ and $L$, respectively. The position of the three readout phases, load (L), read (R) and empty (E) are shown by the white circles in these diagrams. As discussed in the main text both readout techniques rely on a spin-dependent tunnelling process~\cite{elzerman2004}, this can be seen in the data presented in the bottom panels of Fig.~\ref{fig:readout}a and b. Here, the average SET current in time is shown as the read voltage is stepped along the axis shown by the white arrow in the upper panels (equivalent to the detuning axis $\epsilon$ in the main text). At certain read voltages a short current blip can be seen to occur at the start of the read phase indicating the spin-up of the electron qubit. 

Sequential spin readout of electrons on $L$ and $R$ is carried out using the sequence of pulses shown in Fig.~\ref{fig:readout}c. The first phase of this sequence is the read phase of qubit-$L$, after which a pulse is applied to the read out position of qubit-$R$. Since the readout is independent at these detuning positions i.e. exchange is negligible (as shown in Fig.~1 of the main text) the electron remaining on qubit-$L$ during the read phase of qubit-$R$ has no effect on spin readout fidelity. The following four pulses from 3-6 serve to empty and reload electrons from and to the two qubit sites, these may or may not occur depending on the exact experimental protocol, i.e. whether or not a qubit is being prepared with random spins or deterministically with spin-down. Finally, we pulse to position 7 from which we carry out the exchange pulse as described in Fig.~2a of the main text. 

\section{Calculation of lever arms along the detuning axis $\epsilon$}
The lever arm of the gates along the detuning axis $\epsilon$ can be ascertained from the range in read voltage over which tunnelling due to spin-up electrons on the qubits is observed (the so called \textit{spintail}). This range, labelled $\delta_i$ in the lower panels of Fig.~\ref{fig:readout}a and b, is proportional to the Zeeman splitting via the lever arms, $\alpha_\epsilon^L$ and $\alpha_\epsilon^R$ for qubit-$L$ and -$R$ respectively, by
\begin{equation}
\alpha_\epsilon^i\delta_i = \gamma_e B_z,
\end{equation}
\noindent where $\gamma_e{=}28.024$~GHz/T and $B_z$ is the magnetic field. From Fig.~\ref{fig:readout} the lever arms were calculated to be $\alpha_\epsilon^L{=}0.041{\pm}0.004$ and $\alpha_\epsilon^R{=}0.030{\pm}0.003$. The sum of these two lever arms represents scaling of the detuning, $\epsilon$, \textit{between} the qubit-$L$ and -$R$ to the gate voltage $V_{GL}$ along this axis. 

\section{Spin Readout fidelities}
The assignment of a spin-up or -down electron from each SET current trace comprises of two separate parts, (i) electrical readout and (ii) spin-to-charge conversion. 

\begin{table}[h!]
\begin{center}
 \begin{tabular}{||c | c | c||} 
 \hline
 Parameter & Qubit-$L$ & Qubit-$R$ \\ [0.5ex] 
 \hline\hline
 $V_t$~(V) & $0.022{\pm}0.001$ & $0.016{\pm}0.002$ \\ 
 \hline
 $\Delta t$~(ms) & $10.5{\pm}0.1$ & $209.0{\pm}30.0$ \\ 
 \hline
 $F_{\uparrow}$~$(\%)$ & $94.9{\pm}0.8$ & $96.8{\pm}1.6$ \\ 
 \hline
 $F_{\downarrow}$~$(\%)$ & $99.7{\pm}0.2$ & $99.8{\pm}0.2$ \\ 
 \hline
  $\alpha$~$(\%)$ & $99.6{\pm}0.1$ & $99.5{\pm}0.1$ \\ 
 \hline
 $\beta$~$(\%)$ & $98.2{\pm}0.1$ & $99.1{\pm}0.1$ \\ 
 \hline
 $V_{ER}$~$(\%)$ & $94.6{\pm}1.0$ & $96.5{\pm}2.0$ \\ 
 \hline
 $V_{STC}$~$(\%)$ & $97.9{\pm}0.1$ & $98.7{\pm}0.2$ \\
 \hline
 $F_M$~$(\%)$ & $96.2{\pm}1.1$ & $97.6{\pm}2.1$ \\
 \hline
\end{tabular}
\end{center}
\caption{{\bf Experimental parameters for spin readout of qubits $L$ and $R$.}}
\end{table}

\subsubsection{(i) Electrical readout}
The electrical readout involves determining whether a given SET current trace can be assigned as having a `blip' during the read phase i.e. whether during this time the current surpasses a threshold value $I_t$ (see Fig.~\ref{fig:readout}). From a simulation of 10,000 SET traces $50\%$ of which contain a `blip', and with added white Gaussian noise equivalent the signal-to-noise ratio observed in the experiment (average of $SNR{=}17$~dB for readout of both qubits), histograms of peak voltages $V_p$ are generated and shown in Fig.~\ref{fig:fidelity}a and b for $L$ and $R$ respectively. Note the use of peak voltage not current due to the use of a current amplifiera on the drain of the SET charge sensor.

From these histograms the fidelity of assigning either spin-up or -down ($F_\uparrow$ or $F_\downarrow$) to each current trace is calculated using the following set of equations,
\begin{align}
F_\uparrow &= 1-\int_{-\infty}^{V_t}N_\uparrow(V_P)dV_p \\
F_\downarrow &= 1-\int_{V_t}^{\infty}N_\downarrow(V_P)dV_p,
\label{eq:rofids}
\end{align}
\noindent where $V_t$ is the equivalent voltage threshold for $I_t$ after the current amplifier and $N_i$ is the fraction of spin state $i$. The results are shown in Fig.~\ref{fig:fidelity}c and d with the addition of the calculated electrical readout visibility $V_{ER}{=}F_\uparrow{+}F_\downarrow{-}1$. From this we can determine the optimum voltage threshold, $V_t$, where $V_{ER}$ is maximised.

\subsubsection{(ii) Spin-to-charge conversion}
Next we determine the optimum length of time for the read phase of the three level readout sequence. During spin-to-charge conversion errors are introduced from three main sources; $T_1$ relaxation of spin-up electrons; spin-up electrons failing to tunnel to the SET during the designated read time; and spin-down electrons tunnelling to the SET due to thermal excitation~\footnote{Note that the second and third points apply specifically to the spin dependent \textit{unloading} mechanism used at $R$. However, equivalent arguments apply for $L$ in the case of the spin dependent \textit{loading} mechanism.}.

Following from the work of Buch~\cite{buch2013} and Watson~\cite{watsonthesis} we use a rate equation model to determine the optimum readout time, $\Delta t$, based on the probability of a successful assignment of spin-up or -down, $\alpha$ and $\beta$ respectively. As an input to the model the tunnelling times of spin-up out of the qubit site to the SET and spin-down into the qubit site from the SET, $\tau_{\uparrow\text{out}}$ and $\tau_{\downarrow\text{in}}$ are shown in Fig.~\ref{fig:fidelity}e and f. In addition, the spin-down tunnelling time from the qubit site to the SET, $\tau_{\downarrow\text{out}}$ was also measured experimentally to be and $0.61{\pm}0.06$~s and $25{\pm}5$~s for qubit-$L$ and -$R$ respectively. We refer the reader to Ref.~\cite{buch2013} for further details on this model. The readout time, $\Delta t$ vs the fidelities $\alpha$ and $\beta$ are shown in Fig.~\ref{fig:fidelity}e-h. Similarly for the electrical readout, the visibility of spin-to-charge conversion is calculated as $V_{STC}{=}\alpha{+}\beta{-}1$. The optimum readout time is chosen where $V_{STC}$ is maximised. Table 1 gives a summary of the fidelity calculations for both qubits, where the final measurement fidelity is given by, $F_M{=}\left(\alpha F_{\uparrow}{+}\beta F_{\downarrow}\right)/2$. 

\section{Statistical analysis of detuning dependent spin correlations}
The statistical significance of the anti-correlated spin measurements presented in Fig.~2 of the main text can be ascertained using the $\phi$-correlation coefficient (the Pearson correlation coefficient for two binary variables) which is given by,
\begin{equation}
\phi_m = \frac{P_{\uparrow\uparrow}P_{\downarrow \downarrow} - P_{\downarrow\uparrow}P_{\uparrow\downarrow}}{\sqrt{P_{L\uparrow}P_{L\downarrow}P_{R\uparrow}P_{R\downarrow}}}.
\label{eq:phi}
\end{equation}
For perfectly anti-correlated spins $\phi{=}{-}1$, however, given our choice of initial states from Eq.~1 in the main text, where one spin is randomly loaded up or down and the other is deterministically loaded with spin-down, the maximum expected value is $\phi{=}{-}0.25$. We measure an average of $\phi_m{=}{-}0.243\pm0.028$ in the detuning range $0{<}\epsilon{<}2.4$~meV for both $\rho_{\downarrow\uparrow}$ and $\rho_{\uparrow\downarrow}$, see Fig.~\ref{fig:stat}a. The statistical significance of these anti-correlations was deduced from the p-value of $\chi^2{=}\phi_m^2/n$, where $n{=}1000$ is the number of measurement repetitions. A p-value ${\ll}0.01$ is shown in Fig.~\ref{fig:stat}b over the same range of $\epsilon$, demonstrating statistically significant two-electron anti-correlated spins. 

\begin{figure}[h!]
\begin{center}
\includegraphics[width=0.6\textwidth]{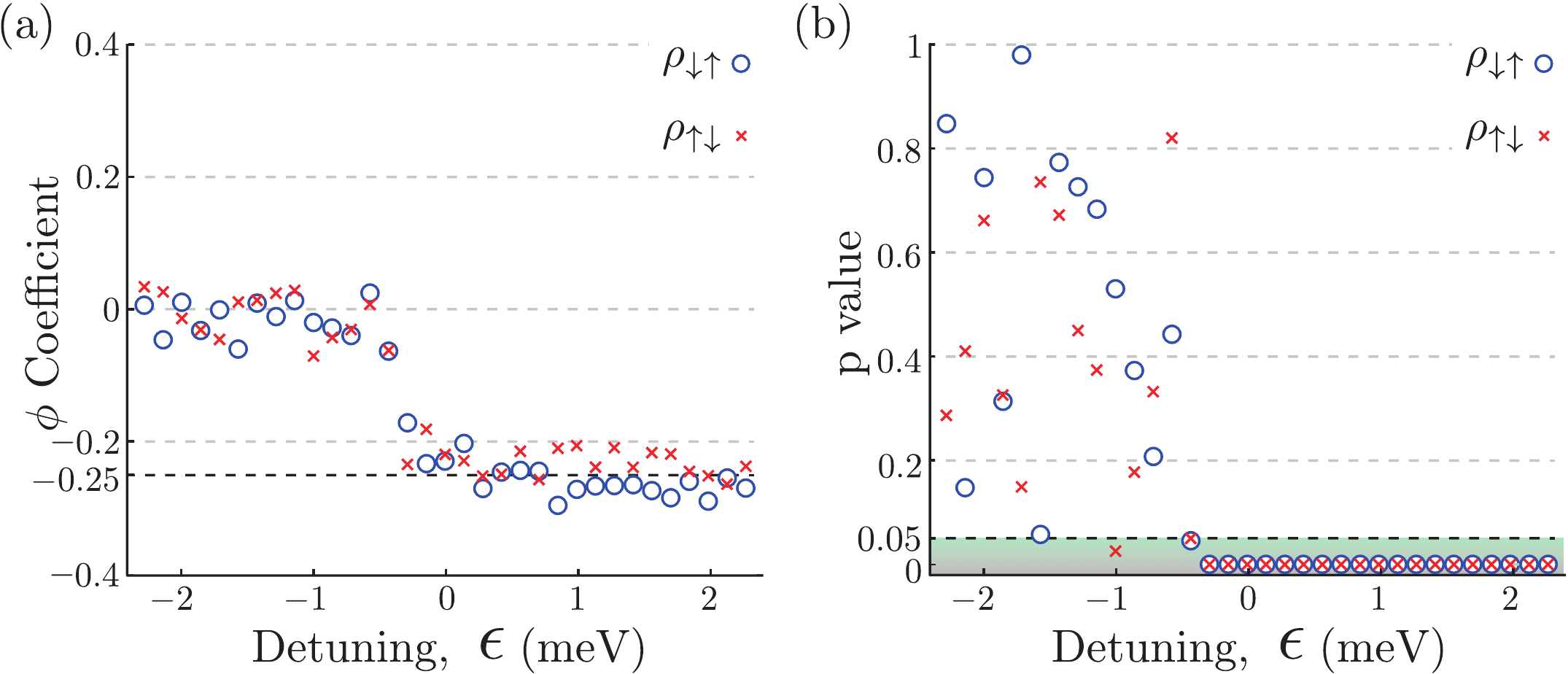}
\end{center}
\caption{{\bf Statistical analysis of detuning dependent spin correlations.} {\bf a}, The $\phi$-correlation coefficient defined in Eq.~\ref{eq:phi} showing the onset of anti-correlated spins as a function of detuning for the initially prepared states $\rho_{\downarrow\uparrow}$ and $\rho_{\uparrow\downarrow}$ given in Eq.~1 of the main text. The expected theoretical value for anti-correlated electron spins is -0.25 (black dashed line). {\bf b}, The p-value of the $\chi^2$ test for this dataset. Data points below a threshold of 0.05 are considered statistically significant.}
\label{fig:stat}
\end{figure}

\section{B-field dependence of single-electron $T_1$ relaxation}
Figure~\ref{fig:T1}a and b shows the dependence of spin-up fraction of each qubit as a function of wait time for different magnetic field strengths. Both qubits show a $B_z^5$ dependence of the $1/T_1$ relaxation rate as shown in Fig.~\ref{fig:T1}c, indicating that the relaxation processes are driven by phonon coupling to the electron spins as previously observed~\cite{morello2010,buch2013,xiao2010}. However, one thing we do have to be careful of is the magnetic field orientation during the measurment. Note that the results for independent readout presented in the main text are shown by square markers in Fig.~\ref{fig:T1}c at $B_z{=}2.5$~T and were obtained with a field orientation of $B_z^{(1)}{||}[1\bar{1}0]$. The magnetic field dependence of  $T_1$ are shown by the red and blue circular markers and were taken during a different cool down with a field orientation $B_z^{(2)}{||}[100]$. Since $T_1$ measurements are known to be highly sensitive to the magnetic field orientation~\cite{Hasegawa1960,Roth1960} the difference in the $T_1$ values observed can be explained by this. 

There are two interesting aspects of the results to note as a result of this different field orientation in Figure~\ref{fig:T1}c. Firstly, we measure the spin relaxation rate of a 2P donor dot in this device to be greater than that for the 1P donor case. This is in contrast to recent theoretical results, which predict a slower relaxation rate for multi-donor qubits in the one-electron case due to its tighter confining potential~\cite{PhysRevLett.113.246406}. Secondly, we see that for the two field orientations $B_z^{(1)}$ and $B_z^{(2)}$ (see inset of Fig.~\ref{fig:T1}c) the 1P relaxation rates are significantly slower than those measured previously ~\cite{morello2010,watson2015}, giving a value of ~22s and ~52s for qubits $L$ and $R$ respectively at B=1.5T. Further work is underway to examine the interplay of the magnetic field and electric field orientation in these Coulomb confined devices.

\begin{figure}[h!]
\begin{center}
\includegraphics[width=1\textwidth]{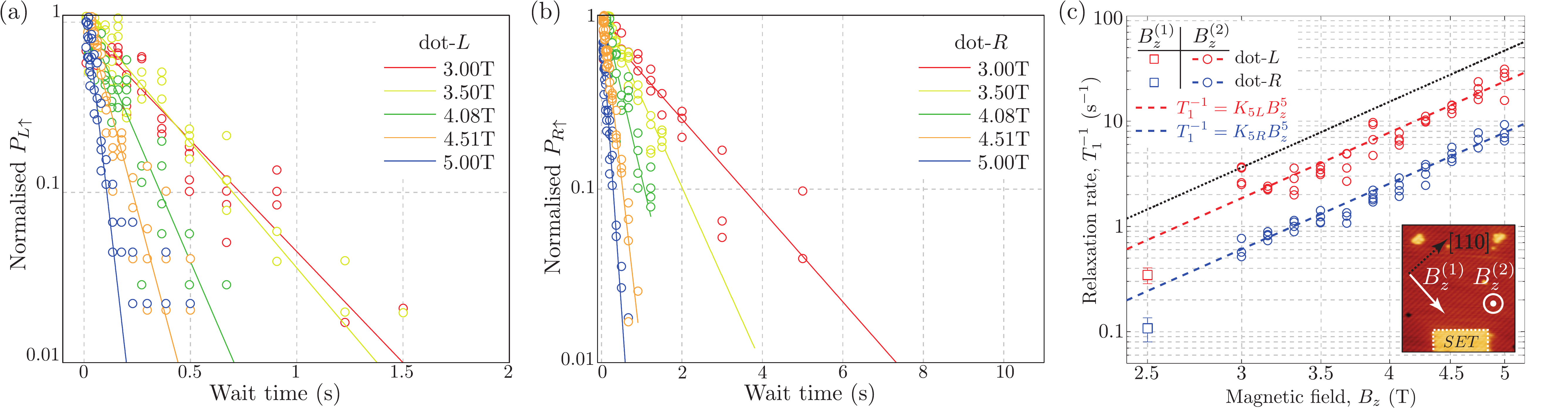}
\end{center}
\caption{{\bf Dependence of donor $T_1$ relaxation times on magnetic field.} {\bf a}, The normalised probability of measuring $\ket{{\uparrow}}$ on $L$ as a function of wait time for different values of magnetic field, $B_z$. Data is plotted with circle markers and fits to exponential decays are given by the solid lines. {\bf b}, An equivalent set of $T_1$ vs. $B_z$ data for $R$. {\bf c}, The $T_1$ relaxation of qubits $L$ and $R$ given in Fig.~1(e and f) of the main text (red and blue squares respectively) was measured at $B_z{=}2.5$~T with the orientation $B_z^{(1)}{||}[1\bar{1}0]$ as shown in the inset. The magnetic field dependence of $T_1$ relaxation was measured for the field orientation $B_z^{(2)}{||}[100]$ (separate cool down), and shown by the red and blue circle markers for qubits $L$ and $R$ respectively. Fits to $T_1^{-1}{=}K_{5i}B_z^5$ are shown by the dashed lines, with the prefactors $K_{5L}{=}0.0060{\pm}0.0010$~s$^{-1}$T$^{-5}$ and $K_{5R}{=}0.0025{\pm}0.0001$~s$^{-1}$T$^{-5}$. Multiple measurements at the same field indicate the spread in $T_1$ value. For comparison the dotted black line shows the result for a single P donor $T_1$ relaxation ($K_5{=}0.015$~s$^{-1}$T$^{-5}$) measured originally by Morello~\cite{morello2010} and confirmed later by Watson~\cite{watson2015} for a field orientation $B_z||[110]$.}
\label{fig:T1}
\end{figure}

\section{Numerical Model of coherent exchange oscillations}
\label{sec:num}
Using the same Hamiltonian given in Eq.~3 of the main text we model coherent exchange oscillations as described in Fig.~3 and~4 of the main text. For this proposed experiment, both electron spins are initialised at a large negative detuning position where the exchange is negligible and subsequently pulsed non-adiabatically into a region where the exchange dominates over hyperfine, that is where $J{>}\Delta B_z$. The spins are allowed to evolve for some time $\tau_w$ before being pulsed back to the initial preparation positions, where spin readout can be carried out. In our simulation we replicate this pulse sequence with the inclusion of detuning noise with a Gaussian distribution~\cite{Hill2009} defined by a standard deviation of 850MHz in detuning energy, based on gate noise measurements of approximately $\delta\epsilon{=}50\mu$V measured in our device. We average over 100 repetitions of the simulation to obtain the final density matrix. In addition to detuning noise, we include a constantly fluctuating Overhauser field which equates to a single electron $T^*_2{=}55$~ns measured in previous electron spin resonance experiments in natural silicon~\cite{morello2010}. Finally, we average over all eight possible donor nuclear spin configurations for the 2P-1P system, giving us an average representation of the nuclear hyperfine interaction. The detuning pulse sequence is simulated for varying $\tau_w$ times, giving rise to the coherent oscillations that can be seen by the green markers in Fig.~3b of the main text. 

\section{Analytical Model of coherent exchange oscillations}
The form of coherent exchange oscillations, as shown in Fig.~4b of the main text, can also be approximated analytically in the following way. Firstly, the oscillation frequency resulting from the exchange interaction, $\nu$, depends on the relative magnitude of the exchange energy $J$ and the difference in magnetic field between the two qubits $\Delta B_z$. For donor systems, $\Delta B_z$ is dominated by the donor nuclear spin orientation resulting in a difference in hyperfine strength between the two qubits, $\delta A$. For a particular nuclear spin configuration the frequency $\nu$ is given by,
\begin{equation}
\nu = \sqrt{J^2+\delta A^2},
\end{equation}
similarly the undamped amplitude of these oscillations $\mathcal{A}$ is given by,
\begin{equation}
\mathcal{A}=\frac{J^2}{J^2+\delta A^2}.
\end{equation}
The amplitude is averaged over all nuclear spin configurations but it is assumed that the frequency is dominated by the most common configuration, which in the 2P-1P case is $\delta A{=}A/2$. 

\begin{figure}[h!]
\begin{center}
\includegraphics[width=0.6\textwidth]{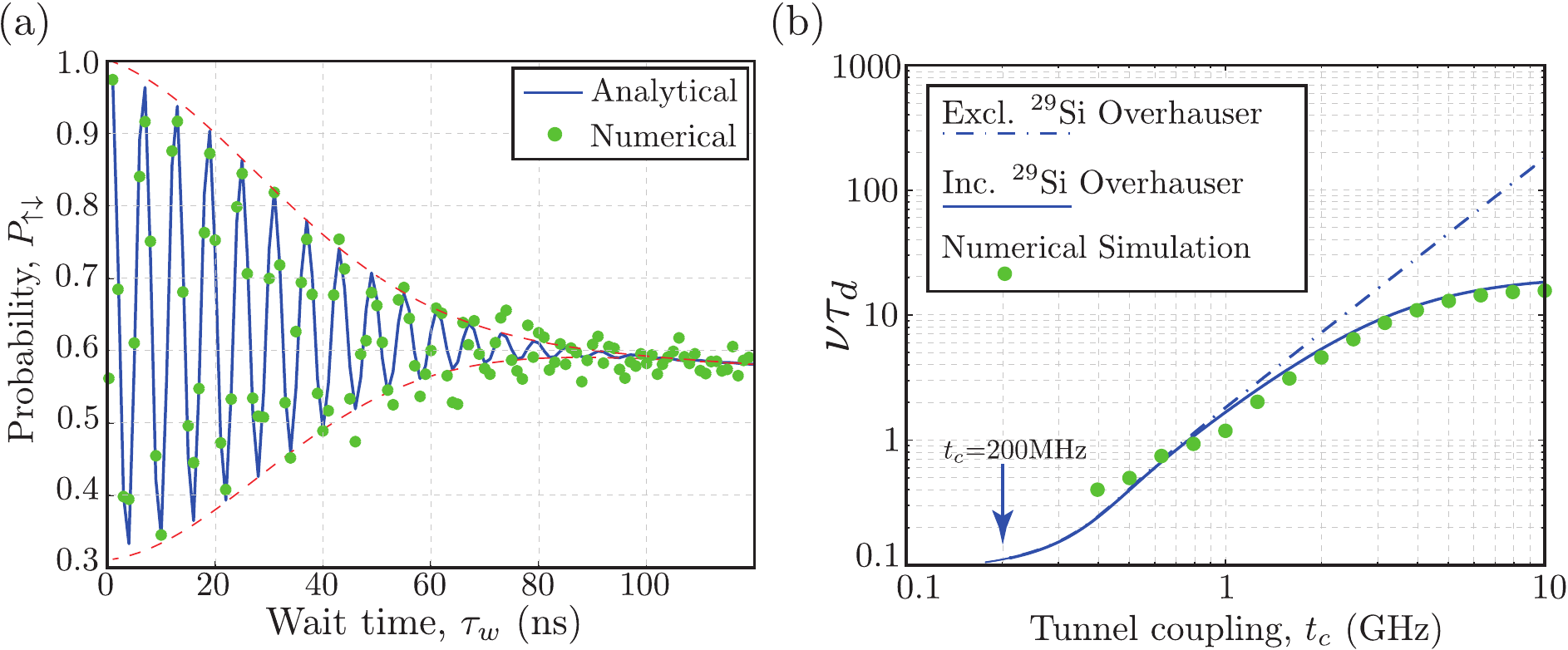}
\end{center}
\caption{{\bf Comparison of analytical and numerical models for coherent exchange oscillations.} {\bf a}, Theoretical prediction of coherent exchange oscillations for a 2P-1P device in natural silicon with tunnel coupling $t_c{=}2.5$~GHz and an applied pulse to $\epsilon{=}{-}25$GHz (circle marker in Fig.4a of main text) for a time $\tau_w$. The two electron state is initialised as $\ket{{\uparrow\downarrow}}$ at a point where the exchange energy is negligible, and subsequently a non-adiabatic detuning pulse is applied to $\epsilon{=}-25$~GHz. We have assumed voltage noise equivalent to 850~MHz along the detuning axis, $\epsilon$  (obtained from measurements) as well as a single electron dephasing time of $T^{*}_2{=}55$~ns measured in previous works~\cite{morello2010}. The results for a numerically simulated full quantum model are shown by the green markers, while the blue line gives the predicted curve based on an analytical expression in Eq.~\ref{eq:theosc}. {\bf b}, Theoretical prediction of $\nu\tau_d$ along the line $\Delta B_z{=}J$ as a function of tunnel coupling for a 2P-1P donor qubit system. Solid (dashed) lines show analytical results including (excluding) the $^{29}$Si Overhauser field, whilst the green markers are results from a numerical simulation.}
\label{fig:comp}
\end{figure}

As in the numerical case, the dephasing is a combination of detuning noise and an Overhauser field. The standard deviation of the detuning noise $\delta\epsilon{=}50\mu$V can be transformed into an exchange frequency noise $\delta \nu$ by considering the minimum and maximum oscillation frequencies given the $\delta\epsilon$,
\begin{equation}
	\delta \nu = \vert \nu(\epsilon+\delta\epsilon,\delta A) - \nu(\epsilon-\delta\epsilon,\delta A) \vert.
\end{equation}
Note here we have also averaged $\nu$ over all possible nuclear spin orientations which give rise to different values of $\delta A$. For a 2P-1P donor dot system $\delta A{=}3A/2$ or $A/2$ with a 1 to 3 ratio. Based on the width, $\delta\nu$, in the frequency domain the resulting dephasing time $\tau_J$ is calculated by converting to the time domain and is given by,
\begin{equation}
	\tau_J=\frac{1}{\pi\delta \nu}.
\end{equation} 	
Note that the decay induced by $\tau_J$ is expected to be Gaussian based on the nature of the noise~\cite{Hill2009}. The total dephasing time $\tau_d$ which includes dephasing from the constantly fluctuating Overhauser field is approximated using the formula,
\begin{equation}
	\frac{1}{\tau_d} \approx \frac{1}{\tau_J} + \frac{1}{T_{2}^*}.
\end{equation}
Finally, the form of the analytical coherent exchange oscillations is given by, 
\begin{equation}
	\frac{1}{2}+ \frac{[\mathcal{A}e^{-(t/\tau_J)^2} \cos(2\pi\nu t) + (1-\mathcal{A})] e^{-t/T_2^*}}{2}.
\label{eq:theosc}
\end{equation}

\noindent Both the characteristic decay time $\tau_d$ and oscillation frequency $\nu$ change as a function of the pulse detuning position $\epsilon$ and tunnel coupling $t_c$, and their product $\tau_d\cdot\nu$ gives an indication of the number of observable oscillations. This is plotted in Fig.~3c of the main text as a function of the tunnel coupling and pulse detuning position. In Fig.~\ref{fig:comp} we shown a comparison of the analytical expressions derived above against a numerical simulation with equivalent parameters as described in previous section. 

\end{widetext}


\section{References}

\begin{thebibliography}{1}

\bibitem{nielsen2010quantum}
Michael~A Nielsen and Isaac~L Chuang.
\newblock {\em Quantum computation and quantum information}.
\newblock Cambridge university press, 2010.

\bibitem{haffner2008}
Hartmut H{\"a}ffner, Christian~F Roos, and Rainer Blatt.
\newblock Quantum computing with trapped ions.
\newblock {\em Physics reports}, 469(4):155--203, 2008.

\bibitem{UniversalLO:OBrien}
Jacques Carolan, Christopher Harrold, Chris Sparrow, Enrique
  Mart{\'\i}n-L{\'o}pez, Nicholas~J. Russell, Joshua~W. Silverstone, Peter~J.
  Shadbolt, Nobuyuki Matsuda, Manabu Oguma, Mikitaka Itoh, Graham~D. Marshall,
  Mark~G. Thompson, Jonathan C.~F. Matthews, Toshikazu Hashimoto, Jeremy~L.
  O{\textquoteright}Brien, and Anthony Laing.
\newblock Universal linear optics.
\newblock {\em Science}, 349(6249):711--716, 2015.

\bibitem{clarke2008}
John Clarke and Frank~K Wilhelm.
\newblock Superconducting quantum bits.
\newblock {\em Nature}, 453(7198):1031--1042, 2008.

\bibitem{petta2005}
J.~R. Petta, A.~C. Johnson, J.~M. Taylor, E.~A. Laird, A.~Yacoby, M.~D. Lukin,
  C.~M. Marcus, M.~P. Hanson, and A.~C. Gossard.
\newblock Coherent manipulation of coupled electron spins in semiconductor
  quantum dots.
\newblock {\em Science}, 309:2180--2184, 2005.

\bibitem{loss1998}
D.~Loss and D.~P. DiVincenzo.
\newblock Quantum computation with quantum dots.
\newblock {\em Phys. Rev. A}, 57:120--126, 1998.

\bibitem{kane1998}
B.~E. Kane.
\newblock A silicon-based nuclear spin quantum computer.
\newblock {\em Nature}, 393:133--137, 1998.

\bibitem{Veldhorst:2015qv}
M.~Veldhorst, C.~H. Yang, J.~C.~C. Hwang, W.~Huang, J.~P. Dehollain, J.~T.
  Muhonen, S.~Simmons, A.~Laucht, F.~E. Hudson, K.~M. Itoh, A.~Morello, and
  A.~S. Dzurak.
\newblock A two-qubit logic gate in silicon.
\newblock {\em Nature}, 526(7573):410--414, 10 2015.

\bibitem{kalra2014}
R.~Kalra, A.~Laucht, C.~D. Hill, and A.~Morello.
\newblock Robust two-qubit gates for donors in silicon controlled by hyperfine
  interactions.
\newblock {\em Phys. Rev. X}, 4:021044, 2014.

\bibitem{nowack2011}
K.~C. Nowack, M.~Shafiei, M.~Laforest, G.~E. D.~K. Prawiroatmodjo, L.~R.
  Schreiber, C.~Recihl, W.~Wegscheider, and L.~M.~K. Vandersypen.
\newblock Single-shot correlations and two-qubit gate of solid-state spins.
\newblock {\em Science}, 333:1269--1272, 2011.

\bibitem{morello2010}
A.~Morello, J.~J. Pla, F.~A. Zwanenburg, K.~W. Chan, K.~Y Tan, H.~Huebl,
  M.~Mottonen, C.~D. Nugroho, C.~Yang, J.~A. {van Donkelaar}, A.~D.~C. Alves,
  D.~N. Jamieson, C.~C. Escott, L.~C.~L. Hollenberg, R.~G. Clark, and A.~S.
  Dzurak.
\newblock Single-shot readout of an electron spin in silicon.
\newblock {\em Nature}, 467:687--691, 2010.

\bibitem{koiller2002}
B.~Koiller, X.~Hu, and S.~{Das Sarma}.
\newblock Exchange in silicon-based quantum computer architecture.
\newblock {\em Phys. Rev. Lett.}, 88:027903, 2002.

\bibitem{Wellard2004}
C~J Wellard, L~C~L Hollenberg, L~M Kettle, and H-S Goan.
\newblock Voltage control of exchange coupling in phosphorus doped silicon.
\newblock {\em Journal of Physics: Condensed Matter}, 16(32):5697, 2004.

\bibitem{Wellard2005}
C.~J. Wellard and L.~C.~L. Hollenberg.
\newblock Donor electron wave functions for phosphorus in silicon: Beyond
  effective-mass theory.
\newblock {\em Phys. Rev. B}, 72:085202, Aug 2005.

\bibitem{Saraiva2015}
A.~L. Saraiva, A.~Barna, M.~J. Calderon, and B.~Koiller.
\newblock Theory of one and two donors in silicon.
\newblock {\em J. Phys.: Condens. Matter}, 27:154208, 2015.

\bibitem{Gamble2015}
John~King Gamble, N.~Tobias Jacobson, Erik Nielsen, Andrew~D. Baczewski,
  Jonathan~E. Moussa, In\`es Monta\~no, and Richard~P. Muller.
\newblock Multivalley effective mass theory simulation of donors in silicon.
\newblock {\em Phys. Rev. B}, 91:235318, Jun 2015.

\bibitem{Rahman2011}
Rajib Rahman, Seung~H Park, Gerhard Klimeck, and Lloyd C~L Hollenberg.
\newblock Stark tuning of the charge states of a two-donor molecule in silicon.
\newblock {\em Nanotechnology}, 22(22):225202, 2011.

\bibitem{Cullis1970}
P.~R. Cullis and J.~R. Marko.
\newblock Determination of the donor pair exchange energy in phosphorus-doped
  silicon.
\newblock {\em Phys. Rev. B}, 1:632--637, Jan 1970.

\bibitem{jamieson2005}
David~Norman Jamieson, CHANGYI Yang, T~Hopf, SM~Hearne, CI~Pakes, S~Prawer,
  M~Mitic, E~Gauja, SE~Andresen, FE~Hudson, et~al.
\newblock Controlled shallow single-ion implantation in silicon using an active
  substrate for sub-20-kev ions.
\newblock {\em Applied Physics Letters}, 86(20):2101, 2005.

\bibitem{simmonsPRL2003}
S.~R. Schofield, N.~J. Curson, M.~Y. Simmons, F.~J. Rue\ss{}, T.~Hallam,
  L.~Oberbeck, and R.~G. Clark.
\newblock Atomically precise placement of single dopants in {Si}.
\newblock {\em Phys. Rev. Lett.}, 91:136104, Sep 2003.

\bibitem{fuechsle2012}
M.~Fuechsle, J.~A. Miwa, S.~Mahapatra, H.~Ryu, S.~Lee, O.~Warschkow, L.~C.~L.
  Hollenberg, G.~Klimeck, and M.~Y. Simmons.
\newblock A single-atom transistor.
\newblock {\em Nature Nanotech.}, 7:242--246, 2012.

\bibitem{Wang2016}
Yu~Wang, Archana Tankasala, Lloyd C~L Hollenberg, Gerhard Klimeck, Michelle~Y
  Simmons, and Rajib Rahman.
\newblock Highly tunable exchange in donor qubits in silicon.
\newblock {\em Npj Quantum Information}, 2:16008, 2016.

\bibitem{buch2013}
H.~Buch, S.~Mahapatra, R.~Rahman, A.~Morello, and M.~Y. Simmons.
\newblock Spin readout and addressability of phosphorous-donor clusters in
  silicon.
\newblock {\em Nature Commun.}, 4:2017, 2013.

\bibitem{weber2014}
B.~Weber, Y.~H.~Matthias Tan, S.~Mahapatra, T.~F. Watson, H.~Ryu, R.~Rahman,
  L.~C.~L. Hollenberg, G.~Kilmeck, and M.~Y. Simmons.
\newblock Spin blockade and exchange in {Coulomb}-confined silicon double
  quantum dots.
\newblock {\em Nature Nanotech.}, 9:430--435, 2014.

\bibitem{Watsone1602811}
Thomas~F. Watson, Bent Weber, Yu-Ling Hsueh, Lloyd C.~L. Hollenberg, Rajib
  Rahman, and Michelle~Y. Simmons.
\newblock Atomically engineered electron spin lifetimes of 30 s in silicon.
\newblock {\em Science Advances}, 3(3), 2017.

\bibitem{watson2015}
T.~F. Watson, B.~Weber, M.~G. House, H.~B\"uch, and M.~Y. Simmons.
\newblock High-fidelity rapid initialization and read-out of an electron spin
  via the single donor ${D}^{-}$ charge state.
\newblock {\em Phys. Rev. Lett.}, 115:166806, Oct 2015.

\bibitem{dirac1926}
P.~A.~M. Dirac.
\newblock On the theory of quantum mechanics.
\newblock {\em Proc. Roy. Soc. A}, 112(762):661--677, 1926.

\bibitem{House:2015rz}
M.~G. House, T.~Kobayashi, B.~Weber, S.~J. Hile, T.~F. Watson, J.~van~der
  Heijden, S.~Rogge, and M.~Y. Simmons.
\newblock Radio frequency measurements of tunnel couplings and singlet-triplet
  spin states in {Si:P} quantum dots.
\newblock {\em Nat Commun}, 6(8848), 2015.

\bibitem{weast1988crc}
Robert~C Weast, Melvin~J Astle, William~H Beyer, et~al.
\newblock {\em CRC handbook of chemistry and physics}, volume~69.
\newblock CRC press Boca Raton, FL, 1988.

\bibitem{Rahman2007}
Rajib Rahman, Cameron~J. Wellard, Forrest~R. Bradbury, Marta Prada, Jared~H.
  Cole, Gerhard Klimeck, and Lloyd C.~L. Hollenberg.
\newblock High precision quantum control of single donor spins in silicon.
\newblock {\em Phys. Rev. Lett.}, 99:036403, Jul 2007.

\bibitem{pla2012}
J.~J. Pla, K.~Y. Tan, J.~P. Dehollain, W.~H. Lim, J.~J.~L. Morton, D.~N.
  Jamieson, A.~S. Dzurak, and A.~Morello.
\newblock A single-atom electron spin qubit in silicon.
\newblock {\em Nature}, 489:541--545, 2012.

\bibitem{Hu2006}
Xuedong Hu and S.~Das~Sarma.
\newblock Charge-fluctuation-induced dephasing of exchange-coupled spin qubits.
\newblock {\em Phys. Rev. Lett.}, 96:100501, Mar 2006.

\bibitem{doi:10.1021/nl3012903}
Bent Weber, Suddhasatta Mahapatra, Thomas~F. Watson, and Michelle~Y. Simmons.
\newblock Engineering independent electrostatic control of atomic-scale (∼4
  nm) silicon double quantum dots.
\newblock {\em Nano Letters}, 12(8):4001--4006, 2012.

\bibitem{koiller2002b}
B.~Koiller, X.~Hu, and S.~{Das Sarma}.
\newblock Strain effects on silicon donor exchange: {Quantum computer
  architecture considerations}.
\newblock {\em Phys. Rev. B.}, 66:115201, 2002.

\bibitem{bussmann2014imaging}
E~Bussmann, M~Rudolph, GS~Subramania, S~Misra, SM~Carr, E~Langlois,
  J~Dominguez, T~Pluym, MP~Lilly, and MS~Carroll.
\newblock Imaging and registration of buried atomic-precision donor devices
  using scanning capacitance microscopy.
\newblock {\em arXiv preprint arXiv:1410.4793}, 2014.

\bibitem{fuhrer2009}
A.~Fuhrer, M.~Fuechsle, T.~C.~G. Reusch, B.~Weber, and M.~Y. Simmons.
\newblock Atomic-scale, all epitaxial in-plane gated donor quantum dot in
  silicon.
\newblock {\em Nano Lett.}, 9:707--710, 2009.

\bibitem{weber2012b}
B.~Weber, S.~Mahapatra, H.~Ryu, S.~Lee, A.~Fuhrer, T.~C.~G. Reusch, D.~L.
  Thompson, W.~C.~T. Lee, G.~Klimeck, L.~C.~L. Hollenberg, and M.~Y. Simmons.
\newblock Ohm's law survives to the atomic scale.
\newblock {\em Science}, 335:64--67, 2012.

\end{thebibliography}

\section{References}
\begin{thebibliography}{2}

\bibitem{hile2015}
S.~J. Hile, M.~G. House, E.~Peretz, J.~Verduijn, D.~Widmann, T.~Kobayashi,
  S.~Rogge, and M.~Y. Simmons.
\newblock Radio frequency reflectometry and charge sensing of a precision
  placed donor in silicon.
\newblock {\em Appl. Phys. Lett.}, 107:093504, 2015.

\bibitem{elzerman2004}
J.~M. Elzerman, R.~Hanson, L.~H. {Willems van Beveren}, B.~Witkamp, L.~M.~K.
  Vandersypen, and L.~P. Kouwenhoven.
\newblock Single-shot read-out of an individual electron spin in a quantum dot.
\newblock {\em Nature}, 430:431--435, 2004.

\bibitem{saraiva2015}
A.~L. Saraiva, A.~Barna, M.~J. Calderon, and B.~Koiller.
\newblock Theory of one and two donors in silicon.
\newblock {\em J. Phys.: Condens. Matter}, 27:154208, 2015.

\bibitem{weber2014}
B.~Weber, Y.~H.~Matthias Tan, S.~Mahapatra, T.~F. Watson, H.~Ryu, R.~Rahman,
  L.~C.~L. Hollenberg, G.~Kilmeck, and M.~Y. Simmons.
\newblock Spin blockade and exchange in {Coulomb}-confined silicon double
  quantum dots.
\newblock {\em Nature Nanotech.}, 9:430--435, 2014.

\bibitem{fuechsle2012}
M.~Fuechsle, J.~A. Miwa, S.~Mahapatra, H.~Ryu, S.~Lee, O.~Warschkow, L.~C.~L.
  Hollenberg, G.~Klimeck, and M.~Y. Simmons.
\newblock A single-atom transistor.
\newblock {\em Nature Nanotech.}, 7:242--246, 2012.

\bibitem{buch2013}
H.~Buch, S.~Mahapatra, R.~Rahman, A.~Morello, and M.~Y. Simmons.
\newblock Spin readout and addressability of phosphorous-donor clusters in
  silicon.
\newblock {\em Nature Commun.}, 4:2017, 2013.

\bibitem{watsonthesis}
T.~F. Watson.
\newblock PhD thesis, School of Physics, University of New South Wales, Sydney,
  Australia, 2015.

\bibitem{Note1}
Note that the second and third points apply specifically to the spin dependent
  \protect \textit {unloading} mechanism used at $R$. However, equivalent
  arguments apply for $L$ in the case of the spin dependent \protect \textit
  {loading} mechanism.

\bibitem{morello2010}
A.~Morello, J.~J. Pla, F.~A. Zwanenburg, K.~W. Chan, K.~Y Tan, H.~Huebl,
  M.~Mottonen, C.~D. Nugroho, C.~Yang, J.~A. {van Donkelaar}, A.~D.~C. Alves,
  D.~N. Jamieson, C.~C. Escott, L.~C.~L. Hollenberg, R.~G. Clark, and A.~S.
  Dzurak.
\newblock Single-shot readout of an electron spin in silicon.
\newblock {\em Nature}, 467:687--691, 2010.

\bibitem{xiao2010}
M.~Xiao, M.~G. House, and H.~W. Jiang.
\newblock Measurement of the spin relaxation time of single electrons in a
  silicon metal-oxide-semiconductor-based quantum dot.
\newblock {\em Phys. Rev. Lett.}, 104:096801, Mar 2010.

\bibitem{Hasegawa1960}
Hiroshi Hasegawa.
\newblock Spin-lattice relaxation of shallow donor states in ge and si through
  a direct phonon process.
\newblock {\em Phys. Rev.}, 118:1523--1534, Jun 1960.

\bibitem{Roth1960}
Laura~M. Roth.
\newblock $g$ factor and donor spin-lattice relaxation for electrons in
  germanium and silicon.
\newblock {\em Phys. Rev.}, 118:1534--1540, Jun 1960.

\bibitem{PhysRevLett.113.246406}
Yu-Ling Hsueh, Holger B\"uch, Yaohua Tan, Yu~Wang, Lloyd C.~L. Hollenberg,
  Gerhard Klimeck, Michelle~Y. Simmons, and Rajib Rahman.
\newblock Spin-lattice relaxation times of single donors and donor clusters in
  silicon.
\newblock {\em Phys. Rev. Lett.}, 113:246406, Dec 2014.

\bibitem{watson2015}
T.~F. Watson, B.~Weber, M.~G. House, H.~B\"uch, and M.~Y. Simmons.
\newblock High-fidelity rapid initialization and read-out of an electron spin
  via the single donor ${D}^{-}$ charge state.
\newblock {\em Phys. Rev. Lett.}, 115:166806, Oct 2015.

\bibitem{Hill2009}
L.~T. Hall, J.~H. Cole, C.~D. Hill, and L.~C.~L. Hollenberg.
\newblock Sensing of fluctuating nanoscale magnetic fields using
  nitrogen-vacancy centers in diamond.
\newblock {\em Phys. Rev. Lett.}, 103:220802, Nov 2009.

\end{thebibliography}
\end{document}